\documentclass[lettersize,journal]{IEEEtran}
\usepackage{amsmath,amsfonts}
\usepackage{algorithmic}
\usepackage{algorithm}
\usepackage{array}
\usepackage[caption=false,font=normalsize,labelfont=sf,textfont=sf]{subfig}
\usepackage{textcomp}
\usepackage{stfloats}
\usepackage{url}
\usepackage{verbatim}
\usepackage{graphicx}
\usepackage{cite}
\usepackage{color}
\usepackage{hyperref}
\usepackage{booktabs} % 导入三线表需要的宏包
\usepackage{tabularray}
\usepackage{multirow}
\usepackage[normalem]{ulem}
\usepackage{xcolor}
\usepackage{ulem}
\usepackage{makecell}
\usepackage{graphicx} 

\newcommand{\toolName}[1]{\textit{ConceptThread}}
\newcommand{\revise}[1]{\textcolor{black}{#1}}
\newcommand{\minorRevision}[1]{\textcolor{black}{#1}}

\begin{document}
%% Paper title.
\title{\toolName{}: Visualizing Threaded Concepts in MOOC Videos}

\author{
Zhiguang Zhou, Li Ye, Lihong Cai, Lei Wang, Yigang Wang, Yongheng Wang, Wei Chen, and Yong Wang
        % <-this % stops a space
\thanks{Zhiguang Zhou, Li Ye, Lihong Cai, Lei Wang, and Yigang Wang are with Hangzhou Dianzi University. E-mail: \{zhgzhou, liye, lihongcai, leiwang, yigang.wang\}@hdu.edu.cn}
\thanks{Yongheng Wang is with Zhejiang Lab. E-mail: wangyh@zhejianglab.com}
\thanks{Wei Chen is with Zhejiang University. E-mail: chenvis@zju.edu.cn}
\thanks{Yong Wang is with Singapore Management University. E-mail: yongwang@smu.edu.sg}
\thanks{(Corresponding author: Yong Wang)}
}

\markboth{IEEE TRANSACTIONS ON VISUALIZATION AND COMPUTER GRAPHICS, VOL. xx, NO. x, JUNE 20xx}%
{Shell \MakeLowercase{\textit{et al.}}: A Sample Article Using IEEEtran.cls for IEEE Journals}

\maketitle
%% Author ORCID IDs should be specified using \authororcid like below inside
%% of the \author command. ORCID IDs can be registered at https://orcid.org/.
%% Include only the 16-digit dashed ID.
%\author{%
  %\authororcid{Josiah S.\ Carberry}{0000-0002-1825-0097},
  %Ed Grimley, and 
  %Martha Stewart
%}

%\authorfooter{
  %% insert punctuation at end of each item
  %\item
  	%Josiah Carberry is with Brown University.
  	%E-mail: jcarberry@example.com
  %\item
  	%Ed Grimley is with Grimley Widgets, Inc.
  	%E-mail: ed.grimley@example.com.

  %\item Martha Stewart is with Martha Stewart Enterprises at Microsoft
  %Research.
  	%E-mail: martha.stewart@example.com.
%}

%% Abstract section.
\begin{abstract}
    Massive Open Online Courses (MOOCs) platforms are becoming increasingly popular in recent years. Online learners need to watch the whole course video on MOOC platforms to learn the underlying new knowledge, which is often tedious and time-consuming due to the lack of a quick overview of the covered knowledge and their structures.
    In this paper, we propose \toolName{}, a visual analytics approach to effectively show the concepts and the relations among them to facilitate effective online learning.
    Specifically, given that the majority of MOOC videos contain slides, we first leverage video processing and speech analysis \revise{techniques}, including shot recognition, speech recognition and topic modeling, to extract core knowledge concepts and construct the hierarchical and temporal relations among them.
    Then, by using a metaphor of thread, we present a novel visualization to intuitively display the concepts based on video sequential flow, and enable learners to perform interactive visual exploration of concepts.
    We conducted \revise{a quantitative study}, two case studies, and a user study to \revise{extensively evaluate} \toolName{}. The results demonstrate the effectiveness and usability of \toolName{} in providing online learners with a quick understanding of the knowledge content of MOOC videos.
\end{abstract}

%% Keywords that describe your work. Will show as 'Index Terms' in journal
%% please capitalize first letter and insert punctuation after last keyword
\vspace{-2mm}

%% A teaser figure can be included as follows

%% Uncomment below to disable the manuscript note
%\renewcommand{\manuscriptnotetxt}{}

%% Copyright space is enabled by default as required by guidelines.
%% It is disabled by the 'review' option or via the following command:
%\nocopyrightspace

%%%%%%%%%%%%%%%%%%%%%%%%%%%%%%%%%%%%%%%%%%%%%%%%%%%%%%%%%%%%%%%%
%%%%%%%%%%%%%%%%%%%%%% LOAD PACKAGES %%%%%%%%%%%%%%%%%%%%%%%%%%%
%%%%%%%%%%%%%%%%%%%%%%%%%%%%%%%%%%%%%%%%%%%%%%%%%%%%%%%%%%%%%%%%

%% Tell graphicx where to find files for figures when calling \includegraphics.
%% Note that due to the \DeclareGraphicsExtensions{} call it is no longer necessary
%% to provide the the path and extension of a graphics file:
%% \includegraphics{diamondrule} is completely sufficient.
\graphicspath{{figs/}{figures/}{pictures/}{images/}{./}{photos/}} % where to search for the images

%% Only used in the template examples. You can remove these lines.
% \usepackage{tabu}                      % only used for the table example
% \usepackage{booktabs}                  % only used for the table example
% \usepackage{lipsum}                    % used to generate placeholder text
% \usepackage{mwe}                       % used to generate placeholder figures

% %% We encourage the use of mathptmx for consistent usage of times font
% %% throughout the proceedings. However, if you encounter conflicts
% %% with other math-related packages, you may want to disable it.
% \usepackage{mathptmx}                  % use matching math font

%%%%%%%%%%%%%%%%%%%%%%%%%%%%%%%%%%%%%%%%%%%%%%%%%%%%%%%%%%%%%%%%
%%%%%%%%%%%%%%%%%%%%%% START OF THE PAPER %%%%%%%%%%%%%%%%%%%%%%
%%%%%%%%%%%%%%%%%%%%%%%%%%%%%%%%%%%%%%%%%%%%%%%%%%%%%%%%%%%%%%%%

\vspace{-1mm}
\section{Introduction}

\begin{figure*}
\centering
\begin{minipage}{1\textwidth}
    \centerline{\includegraphics[width=1\textwidth]{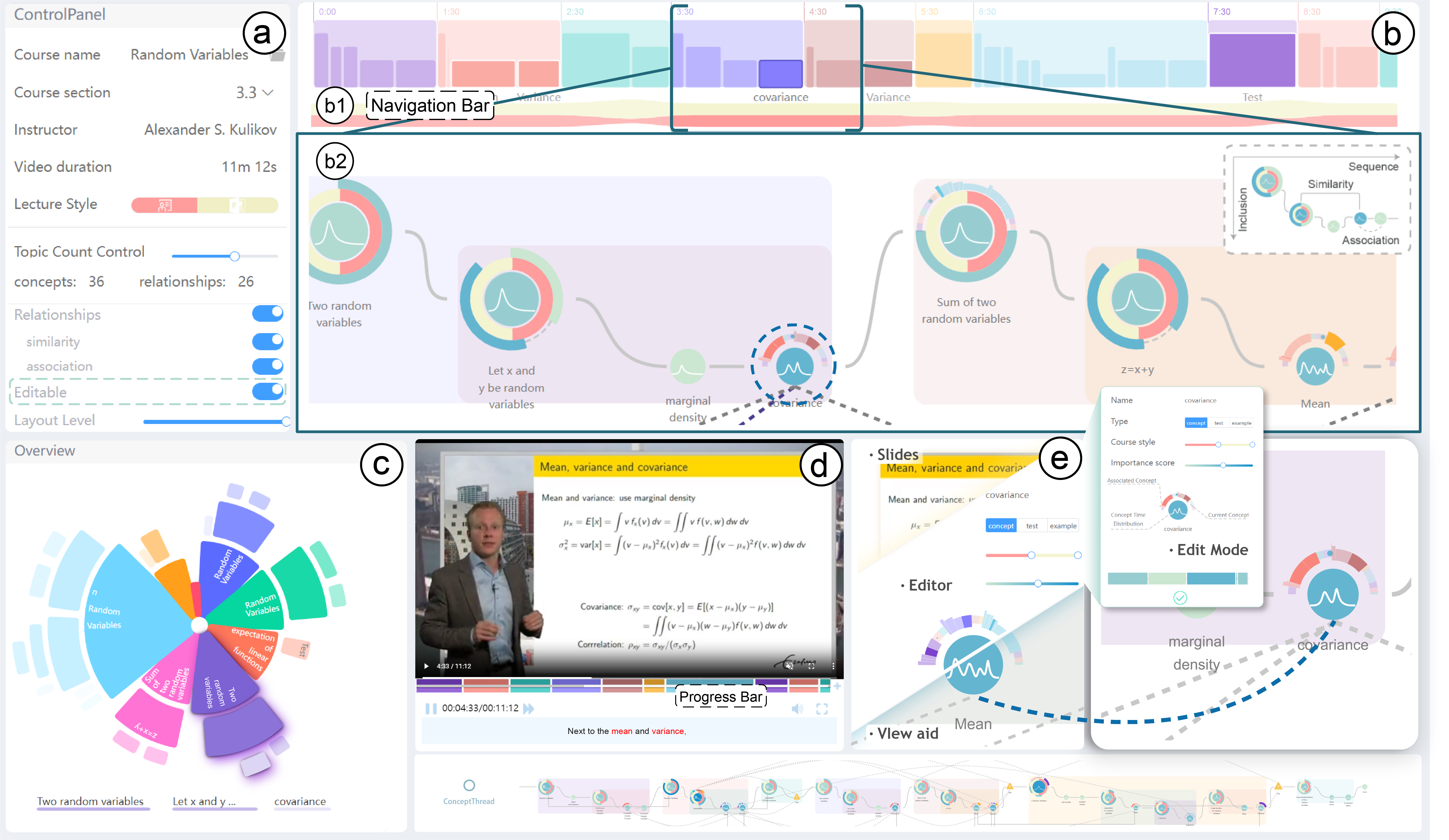}}
\end{minipage}
\caption{
\minorRevision{Exploration of Coursera course videos from \textit{Econometrics: Random Variables} on random variables, with specific information and system configuration displayed in (a). The course's main structure is presented in (c) overview, further divided into layers in (b) to view detailed concepts and their relationships, helping users maximize insight without browsing the video. The corresponding (d) raw data (video) and (e) support panel can further facilitate cognition.}
}

\label{fig:teaser}
\vspace{-3mm}
\end{figure*}

\IEEEPARstart{M}{assive} Open Online Courses (MOOCs) are open-access course platforms, e.g., Coursera, edX, and Udacity, and have attracted increasing attention in the field of education in recent years. As the main 
teaching and \revise{learning content in MOOCs~\cite{breslow2013studying,seaton2014does,ally2004foundations}}, \revise{course videos} 
% \wy{pls update all the terms in the paper.}
provide learners with visual and audio information that helps them construct impressions, relations, and logical structures of knowledge in the brain, ultimately leading to knowledge obtaining and application. However, \revise{watching the full videos} is time- and labor-consuming, and the vast amount of complex knowledge
% in MOOCs 
\revise{can result in overlooking} or misunderstanding concepts.
% missed or misunderstood concepts. 
Moreover, relying on video solely may hinder \revise{learners from effectively recognizing}
the connections between learned knowledge, making it challenging to retain long-term knowledge~\cite{anderson1998online}. Therefore, constructing a comprehensive knowledge exploration system and providing learners with a quick overview of knowledge structures and logical relations is significant in MOOC learning, and is even crucial for further knowledge system construction and learning/thinking method formation~\cite{ally2004foundations}.

%Numerous statistical studies have analyzed the content of MOOC videos and provided valuable insights for learning [5,6,8,9]. Huang et al. [5] proposed a method to visualize video content using keyword clouds, allowing learners to quickly grasp learning concepts by combining interactions and script information. Liu et al. [6]developed ConceptScape, a system that uses concept maps to help users locate video segments related to specific concepts, enabling them to explore the video content and pinpoint the time point of a concept video. Recent empirical studies on MOOC videos have examined which factors in video production affect student engagement, leading to general recommendations [7,11]. However, existing research advocates for manual interactive construction of concept maps through crowdsourcing [6], which is time-consuming and generates concept maps that are not suitable for linear representations based on videos [10], making them non-scalable and non-adaptive. Alternatively, analyzing user behavior data for video content is associated with a high threshold [80]. Therefore, developing knowledge externalization strategies that align with the logical system of the course and are expressed in MOOC videos is crucial for learners to improve their learning efficiency.
Numerous studies have analyzed the content of MOOCs videos and provided valuable insights for learning methods ~\cite{huang2017videomark,liu2018conceptscape,wu2018multimodal, guo2014video, chorianopoulos2013usability}.
% Visualizing knowledge concepts is emerging in recent years to help learners obtain knowledge. 
Visualizing knowledge concepts aims to help learners obtain knowledge \revise{and has become an emerging research topic in recent years}. For example,
Huang et al.~\cite{huang2017videomark} proposed a method to visualize video content by keyword clouds to allow learners to quickly grasp knowledge concepts by combining interactions and script information. Liu et al.~\cite{liu2018conceptscape} developed ConceptScape, a system that provides concept maps to help users locate video segments based on specific concepts, enabling users to pinpoint the time point of concepts. However, existing studies advocate constructing concept maps through crowdsourcing~\cite{liu2018conceptscape}, which is time-consuming. The generated concept maps may not be suitable to represent videos~\cite{novak1984learning}, making them non-scalable and non-adaptive for analysis. Therefore, based on the logical relations of knowledge in videos, developing concept maps, regarded as a specific knowledge externalization strategy, is crucial for learners to improve their learning efficiency.

%After close communication and discussion with experts in the field, we reflected on and summarized the challenges of designing a visualization tool for MOOCs, which consist of three main aspects. Firstly, the multi-modal format of MOOCs videos contains a vast amount of knowledge, but effectively extracting and utilizing this knowledge is difficult (CH1). Secondly, there are various forms of connections between knowledge, and distinguishing and establishing logical relationships among them is challenging (CH2). Finally, externalizing the knowledge structure requires visual representation, so that learners can not only observe the composition of the course's internal knowledge from a macro perspective, but also quickly focus on specific knowledge points and provide good cognitive assistance, which requires excellent visual design (CH3).
After
% close communications and discussions 
\revise{extensive interviews}
with domain experts, we reflected on and summarized the challenges of designing a visualization tool for MOOC videos, which consist of three main aspects. Firstly, extracting and utilizing the knowledge from MOOC videos is difficult due to the existing large-scale multi-modal information and abundant knowledge (CH1). Secondly, distinguishing and establishing logical relationships among the knowledge is challenging, because there are various forms of connections between knowledge in MOOC videos (CH2). 
Finally, \revise{the learning process is phased, and coordinating views to support knowledge dissemination can be challenging. According to Haring et al.'s Instructional Hierarchy (IH) four-stage learning model~\cite{haring1978fourth}, 
the first stage of learning is \textit{Acquisition}, which aims to grasp the broad macro-structure of knowledge, followed by a deeper \textit{Fluency} stage for precise understanding.} Therefore, the system should enable learners to see the course's internal knowledge composition from a macro perspective \revise{while allowing} quick focus on specific concepts and \revise{offering cognitive support} (CH3).
% \wy{The last challenge seems a bit vague to me.}

%Drawing on the MOOC video education methodology [7] and in response to the aforementioned challenges, we introduce ConceptThread: an intuitive, exploratory, and information-rich visual analysis system designed to help learners effectively externalize their understanding of MOOC videos. ConceptThread is grounded in the notion that visualizing the flow of knowledge within MOOC videos through a narrative lens enhances learners' retention of concepts and their interrelationships. To develop ConceptThread, we first extracted the concept expressions from MOOC videos, classified and evaluated important metadata, and then extracted the main concepts through a set of video processing and speech analysis methods. Secondly, based on different logical relationships, we identified four types of relationships that could provide learners with clear insights and further refined them using strategies such as topic modeling and HITF (human in the loop). Finally, we linearly encoded the concepts horizontally and provided intuitive relationship representations, helping learners to learn in detail according to the cognitive path of knowledge, and implemented a conceptual thread metaphor based on video sequence streaming. We also provided an overview of the course content to help learners grasp the course from a macro level. The case study, user research, and expert interviews demonstrate the effectiveness and usability of ConceptThread in providing online learners with a quick understanding of MOOC video knowledge content.
Building upon the MOOC video education methodology~\cite{guo2014video}, we introduce \toolName{}, an intuitive, exploratory, and information-rich visual analytics \revise{approach} to help learners understand MOOC videos. 
% Previous research has indicated that utilizing a narrative structure as a carrier can enhance reading, comprehension, and memorization~\cite{prins2017tell,van1979cognitive}. 
% \wy{what is a carrier here?}
\revise{Previous research has indicated that showing the knowledge structure in a narrative manner can enhance reading, comprehension, and memorization~\cite{prins2017tell,van1979cognitive}.}
% \wy{pls check if this is what you want to say.}
Therefore,
as shown in Fig.~\ref{fig:teaser}, \toolName{} \revise{follows the idea and visualizes the flow of knowledge within} MOOC videos to enhance learners' retention of course concepts and their interrelationships. 
% \wy{Pls confirm if I changed your idea here.}
To develop \toolName{}, we first extracted the \revise{essential element} from MOOC videos,
% \wy{what is concept expressions?}
classified and evaluated important metadata of the concepts, and further extracted the main concepts through a set of video processing and speech analysis methods. Then, based on different logical relationships among concepts, we identified and refined four types of relationships to convey clear insights and logical structures of concepts. Finally, we encoded the concepts horizontally and provided intuitive relationship representations. To assist learners in obtaining knowledge insights expediently, \revise{we employ a thread metaphor and propose a novel flow-based visualization to show both the relation and temporal evolution of concepts in a MOOC video, where a compact radial glyph is present to show the details of each concept.}
% we implemented a thread metaphor based on video sequence flow, accompanied by a set of novel visual design encodings for nodes and edges within \toolName{}.
% \wy{The major novelty of the paper, i.e., the designs, is not clearly highlighted here.}
In addition, we provided an overview of the MOOC video content to help learners grasp the macro-level insights and an integrated video player to facilitate learners' review of the original knowledge content. 
% \wy{What is the point of this sentence?}
Our contributions are summarized as follows:
\begin{itemize}
	\setlength{\itemsep}{0pt}
	\setlength{\parsep}{0pt}
	\setlength{\parskip}{0pt}
  \item \revise{We propose \toolName{}, an interactive visual analytics system to help online learners quickly explore and learn the knowledge concepts in MOOC videos, where a novel flow-based visualization is developed using a thread metaphor to visualize the relationship and temporal evolution of different concepts within a MOOC video.
  }

  % \item \revise{We introduce a set of \textit{visual designs}, which enables the dissemination of MOOC knowledge under video sequential flow by a thread metaphor, and encompasses novel visual representations for concepts and relationships to accommodate the specifications of MOOC videos.}

  \item \revise{We conduct extensive \textit{evaluations} to evaluate the effectiveness and usability of \toolName{}. The results from quantitative studies, case studies, and user studies show that \toolName{} can effectively help online learners conveniently and quickly understand MOOC videos.
  } 
\end{itemize}

% \wy{Reach here.}

\section{RELATED WORK}
\revise{The related work of this paper can be categorized into three groups: MOOC data analysis, video visualization and concept map construction.}
% \textit{MOOC data analysis}, \textit{video visualization} and \textit{concept map construction}.

% \wy{What is the relation of the three subsections here? Subsection B is kind of a subset of Subsection A?}

\vspace{-1mm}
\subsection{\revise{Visual Analytics of MOOC Data}}
\revise{Various MOOC platforms have become increasingly popular for students to conduct online learning in the past decades. Accordingly, different types of MOOC data have been collected, and researchers have attempted to analyze them to enhance the student learning effectiveness~\cite{harasim2000shift,zhao2018flexible} and teachers' teaching designs~\cite{shi2015vismooc,fu2016visual,xia2020seqdynamics}.
According to the study by Kui et al.~\cite{kui2022survey},
existing work on MOOC data analysis can be classified into three groups:
\textit{learning behavior analysis}, \textit{
student interaction analysis
}, and \textit{learning content analysis}.}

\revise{\textit{Learning behavior analysis} explores the relationship between student behaviors and academic outcomes. It is beneficial for teachers to make appropriate learning interventions and improve curriculum designs~\cite {chen2018viseq,xia2020qlens,tsung2022blocklens}. For instance, VisMOOC~\cite{shi2015vismooc} utilizes the seek graph and event graph to illustrate video jump patterns, aiding instructors in understanding student behavior.
PeakVizor~\cite{chen2015peakvizor} focuses on the peaks of MOOC video clickstream data and presents a set of visual designs to show spatio-temporal information about the peaks and the correlation between different learner groups and peaks.
\textit{Student interaction analysis} is an effective way for students to seek help and improve student relationships~\cite{sung2016topin,wu2016networkseer,fu2016visual}. Zheng et al.~\cite{zheng2018measuring} introduced a system for studying group knowledge elaboration during online discussions.
Fu et al.~\cite{fu2018visforum} developed VisForum, a visual analytics system to help MOOC platform owners and instructors analyze the interactions among students, instructors and teaching assistants. }

\textit{Learning content analysis} focuses on analyzing course content and can be categorized into three levels of analysis~\cite{shih2005content}:
object-level, event-level, and conclusion-level. 
1) Object-level analysis analyzes the core concepts in online courses to index, retrieve, and browse the course content.
For example, Chen et al.~\cite{chen2010automatic} extracted key phrases and keywords to get a general idea of the course from course audio channels based on script and slide text. 
Balasubramanian et al.~\cite{balasubramanian2016multimodal} proposed a metadata extraction method to retrieve the course content by utilizing the features from the visual and audio channels of the video. 
2) Event-level analysis can identify patterns between concepts as well as during transitions to analyze their intrinsic connections.
% \wy{For what analysis???} 
For example, Zhang et al.~\cite{zhang2019scaffomapping} enabled users to manually create relationships between concepts by providing frequency and timestamped representations of keywords. 
Also, some studies~\cite{guo2014video,chen2015effects} aided learners in selecting their preferred teaching style by analyzing various teaching methods in educational videos. 
3) Conclusion-level analysis provides the summarization of the courses. 
For instance, Zhao et al.~\cite{zhao2017novel} dynamically demonstrated the segmentation of topics and teachers' participation in the teaching process by segmenting slides and voice information. 
Video retrieval techniques~\cite{yang2014content,zhao2016new} are also applied to help learners navigate across the relevant parts in MOOC videos.

\revise{Our work belongs to \textit{learning content analysis}.
Different from prior work on analyzing learning content, our work combines the course content at all three levels.
We furnish representations for various concept attributes to facilitate object-level scrutiny while also displaying diverse relationships between concepts at the event-level and illustrating course styles. Finally, we present a series of collaborative views summarizing the course content at the conclusion-level,
% \wy{pls clarify how we combine the content of all the three levels }
% we first analyzed and categorized the metadata at each level in the course, and finally extracted the most critical metadata based on user feedback, 
which helps learners gain a deep understanding of the course video.}

\vspace{-1mm}
\subsection{\revise{Video Visualization}}
\revise{Video visualization aims to use visualization to provide users with a quick overview of the video contents, helping users quickly explore and analyze data stories in videos.}

\revise{Previous studies have explored video visualization techniques for various application domains, including sports~\cite{wang2021tac,legg2012matchpad}, online learning~\cite{zhang2022towards,huang2017videomark}, and healthcare~\cite{preim2020survey}. 
Various visualization approaches have been developed to provide valuable insights into the video data of different application domains. 
For example, for sports video analysis, visualization techniques such as high-dimensional data visualization, time-series visualization, and glyph visualization have been widely used to visualize sports videos~\cite{du2021survey}.
For online learning, the visualization of MOOC videos is an emerging research topic~\cite{kui2022survey} and prior
% Notable 
studies have explored different visualization approaches to
% enhance the analysis and understanding of 
analyze
MOOC videos. 
Specifically, Huang et al.~\cite{huang2017videomark} presented an approach that allows learners to quickly view learning concepts in videos using a keyword cloud.
Liu et al.\cite{liu2018conceptscape} employed a node-link graph to facilitate the identification of video segments corresponding to specific concepts.
Also, Davis et al.\cite{davis2016gauging} proposed an arc diagram to visualize the learning path and sequence of video viewing.
Zeng et al.~\cite{zeng2022gesturelens} introduced GestureLens, an interactive visual analytics system designed to facilitate gesture-based and content-based exploration of gesture usage in presentation videos., which helps users explore the association between speech content and gestures through a set of intuitive glyph designs, thus assisting professional public speaking coaches in improving their gesture training for students.}

\revise{In comparison to other visualization approaches, traditional video visualization methods in online learning can be simplistic, Ignoring time series and potentially overlooking hidden patterns and meaningful relationships within course data.
Building upon insights from previous methods, our work explores diverse design alternatives and employs intuitive visual designs to enrich the visualization of video elements.
Our approach not only presents concepts in a sequential flow within videos but also enables learners to engage in interactive visual exploration of the interrelationships of these concepts.}

\vspace{-1mm}
\subsection{\revise{Concept Map Construction}}
\revise{Concept maps are widely recognized as an effective tool for facilitating learning in the field of education~\cite{novak1990concept,nesbit2006learning}. Therefore, how to construct a concept map has become a long-standing research topic in the field of data visualization.
The key to constructing a concept map lies in the extraction of concepts and their relationships.}

\revise{Yadav et al.~\cite{yadav2016vizig} utilized a deep convolutional neural network to automatically identify and classify various anchor points in educational videos, such as graphs, tables, equations and code snippets. 
These anchors are then integrated with a topic list to form the simplest concept map representation for the non-linear navigation of videos.
ConceptScape~\cite{liu2018conceptscape} leveraged crowdsourcing techniques to extract concepts and the relationships among them, forming concept maps that can reflect the connections between concepts.
To enhance the connections between videos, Schwab et al. proposed booc.io~\cite{schwab2016booc}, an interactive visual analytics platform to support dynamic nonlinear learning planning
using hierarchical concept maps.
These methods employ graph-based representations to organize educational materials from various sources, fostering self-regulated learning~\cite{zimmerman2001self}.}

Existing research on concept map construction often leverages deep learning and crowdsourcing techniques to build concept maps. It requires large datasets and much manual effort, making existing approaches often time- and labor-intensive. Also, existing studies primarily focus on graph-based visual designs
and do not support showing the knowledge structure in a narrative manner.
Different from them,
our approach automates the process of concept map construction and a novel flow-like visualization design using a thread metaphor is presented to represent concepts and their relationships, which intuitively displays how the MOOC course instructors teach a class in a MOOC video.

\section{REQUIREMENT ANALYSIS}

% In this section, we summarize the analysis tasks and identify the main design requirements by interviewing relevant personnel and reviewing previous literature. 
\revise{In this section, we conducted a preliminary survey to investigate the important elements required for concept map construction. We then identified specific design requirements through interviews with domain experts.}

\subsection{\revise{Survey on Constructing Concept Maps in MOOC videos}}
\label{sec:task_analysis}
\revise{We first identified essential elements of MOOC videos. Our goal was to build a concept map expressing course content. Existing concept map studies~\cite{novak2006theory,barrachina2015design} posit that they consist of basic nodes and the associated edges. Nodes contain \textit{concepts} themselves and \textit{propositions} composed of many concepts, and edges represent the \textit{relationship} between concepts. Such relationships still exist in MOOC videos, where \textit{concepts} represent small knowledge units, and \textit{propositions} group them. Notably, courses typically consist of a few root propositions (also known as the root topics) at the highest level in the time series to guide the structure of the course, and further divide it into finer concepts and relationships, this feature also influences our data processing workflow.}

\revise{After obtaining the most basic elements from the research of concept maps, we looked for elements from the learner's perspective that they focus on when studying course videos.
We conducted interviews with 10 college students who had been using MOOCs for more than three years (4 in computer science, 3 in finance, and 3 in art design majors). 
To assist learners in understanding the pattern of knowledge dissemination during teaching, we delved into the cognitive aspects of concepts,
McCarthy's 4-MAT Cycle of Learning~\cite{AboutLearning}, with its phases of \textit{Preparation}, \textit{Demonstration}, \textit{Application}, and \textit{Integration}, is closely aligned with MOOC concept cognition. 
We used this framework to guide our interviews and gather insights into what learners primarily focus on during each phase.
Each was asked questions about MOOC learning, such as (1) What elements do they consider important for facilitating learning and memory during the process of learning MOOCs? (2) What elements do they consider desirable but not explicitly represented in MOOC teaching? 
Throughout the interview process, they were free to browse courses on various MOOC platforms to facilitate their responses.}

\revise{We identified the elements that students most frequently mentioned during these stages.
In the \textit{Preparation} phase, they often need to link concepts to existing knowledge by understanding the \textit{relationships} between knowledge, and the \textit{concept time distribution} were also considered an important element to aid in repetitive understanding.
In the \textit{Demonstration} phase, they need to learn the concepts in detail, as \textit{slides} are considered the most important means of facilitating understanding since they contain the most general content describing the concepts.
They then progress to the \textit{Application} phase to understand how concepts are applied. In this phase, \textit{examples} and \textit{tests} were the most frequently mentioned elements. Students generally considered these as the most common means teachers used to facilitate the application of concepts.
Finally, they entered the \textit{Integration} phase, where learners transform knowledge into their own by completing homework.
However, homework is typically included as separate modules within MOOCs and is not usually presented in the videos. Therefore, we did not take them into consideration.}

\revise{Additionally, we conducted a comprehensive review of previous visual analytics systems and performed a statistical analysis of the findings. In addition to the mentioned basic elements, elements like \textit{course style} (teaching aids used to deliver the course, e.g., slides and drawing board)~\cite{chen2015effects,chen2015peakvizor,chorianopoulos2013usability} and \textit{key concepts}~\cite{shi2015vismooc,zhang2019scaffomapping,zhao2017novel} garnered significant attention from learners in previous applications.}

\revise{Furthermore, we have structured \textit{relationships} to enhance learners' understanding. Donald~\cite{galambos1986knowledge} categorized the relationships between concepts into three types: associative, functional, and structural, which are further mapped to \textit{\uline{Sequence}}, \textit{\uline{Association}}, \textit{\uline{Similarity}}, and \textit{\uline{Inclusion}} to better express the relationships between concepts in MOOC videos. Any relationship between concepts and propositions can be expressed through these four types of relationships.}

\revise{Based upon this, the concept map in MOOC videos should consist of \textit{concepts} and their \textit{relationships}, where concepts can be further categorized into their \textit{content} (slides, tests, and examples) and \textit{attributes} (course style, key concepts, and concept time distribution).}

% \begin{figure}[b]
% 	\centering

% 	\includegraphics[width=3.5in]{figs/Fig3-revised.png}

% 	\caption{Results from a pilot study. Course metadata are categorized based on their \revise{perceptual range} and \revise{analysis degree} to form concept content. Bold data constitute the data structure of the system. Four types of relationships between concepts are listed below.}
% 	\label{Fig2}
% \end{figure}

% Collaborated with the analysis result and suggestions of domain experts, five tasks (T1-T5) of this work are summarized and refined as follows.

% Afterwards, we collaborated with MOOC instructors from Coursera and edX platforms, along with an educational analyst who had a background in MOOC research, to discuss the results of our previous student survey. Specifically, we focused on the conceptual cognition issues that students may encounter when learning MOOC videos. Meanwhile, experts provided insights into the difficulties they encountered in online education and popular analysis tasks in their respective fields. They also helped us refine questions and suggested additional tasks to analyze potential teaching patterns that learners might not generally be aware of. Through several rounds of meetings, we gained a better understanding of the problems and summarized the main task requirements based on expert feedback and knowledge gained from past experiences. Throughout the task analysis and prototype design stages, we followed a user-centered design process as proposed by \cite{munzner2009nested}.

\subsection{Design Requirements}

This paper aims to help learners gain a better understanding of MOOC videos. We worked closely with two experts~(E1 \& E2) in online learning to extract the detailed requirements and collect their feedback.
E1 is an instructor who has been offering online lecture videos on Coursera and edX platforms for 3 years and E2 is an educational analyst who has a background in MOOC video research for 10 years.
We conducted discussions with them through \minorRevision{VooV Meeting}~\footnote{\minorRevision{VooV Meeting: A cloud-based video conferencing platform.~\url{https://voovmeeting.com/index.html}}}
% \wy{Why adding this? Perhaps we can simply add a footnote here to show what it is.}
and face-to-face discussions.
We begin by presenting prior findings from interviews with college students. Specifically, we focused on the cognition issues that learners may encounter when learning MOOC videos. Meanwhile, experts provided insights into the difficulties and popular analysis tasks they encountered in online education. They also helped us refine requirements and suggested additional tasks to analyze potential teaching patterns that learners might not generally be aware of. Through several rounds of meetings, we gained a better understanding of the problems and summarized six design requirements (R1-R6) based on expert feedback and knowledge gained from past experiences.

\textbf{\revise{R1. Show the importance of each concept.}}  Identifying important concepts is essential for learners to improve learning efficiency and comprehension of course content. Both experts agreed that the frequency of concept occurrence (in language or related materials), duration, and correlation with other concepts can indicate the importance of concepts, which need to be explicitly presented to learners.

\textbf{\revise{R2. Reveal the temporal evolution of concepts.}} 
Understanding the temporal evolution of concepts
is crucial for learners to gain a quick overview of course content, enabling learners to plan their learning process better.
Also, by checking the frequency of different concepts and their detailed repetitions, learners can easily know which concepts are relatively more important and when they are introduced or highlighted throughout the whole MOOC video.

\textbf{\revise{R3. Describe the relationship between concepts and propositions.}}
According to E1's insights, during his courses, he will explicitly highlight the relationships between concepts to assist learners in building a better knowledge structure. He also emphasized that there are different kinds of relationships, such as causation, deduction, analogy, and other methods of knowledge elaboration. Clearly demonstrating these relationships will help learners intuitively grasp the inherent connections between concepts.

\revise{\textbf{R4. Outline the organization of the course content.} 
E2 commented, ``It is necessary to analyze and show the course's organizational form, including the hierarchical structures among different concepts and the sequential order of introducing them in a MOOC video. A clear organizational form can not only accelerate learners' learning progress but also deepen their understanding of the course."
It can allow learners to flexibly decide whether they will dive into the details of specific concepts according to their preference, which significantly improves the learning experience and learning efficiency.}

\textbf{\revise{R5. Provide collaborative views to support each stage of learning.}} 
Experts have suggested that learning is multi-stage, and our system should support this learning cycle to help users understand and digest course content, thus promoting efficient learning. After further discussion, experts indicated that the IH four-step learning model~\cite{haring1978fourth} proposed by Haring et al. is well suited for MOOC videos, where learners first acquire knowledge, then achieve fluency and generalization in the use of the acquired knowledge, and finally adapt the knowledge through extensive application.

\revise{\textbf{R6. Enable users to refine automatically generated visual results.} 
After several rounds of discussions with experts on the prototype system, the experts concurred that keeping human-in-the-loop method is an effective strategy for optimizing or correcting the results if the data processing does not perform well. This not only corrects system errors for subsequent learners but also deepens the learners' retention of knowledge. Therefore, options should be provided for users to edit, add/remove concepts, and adjust other metadata.}

\section{System Overview}
\begin{figure}[tb]
	\centering
	\includegraphics[width=\linewidth]{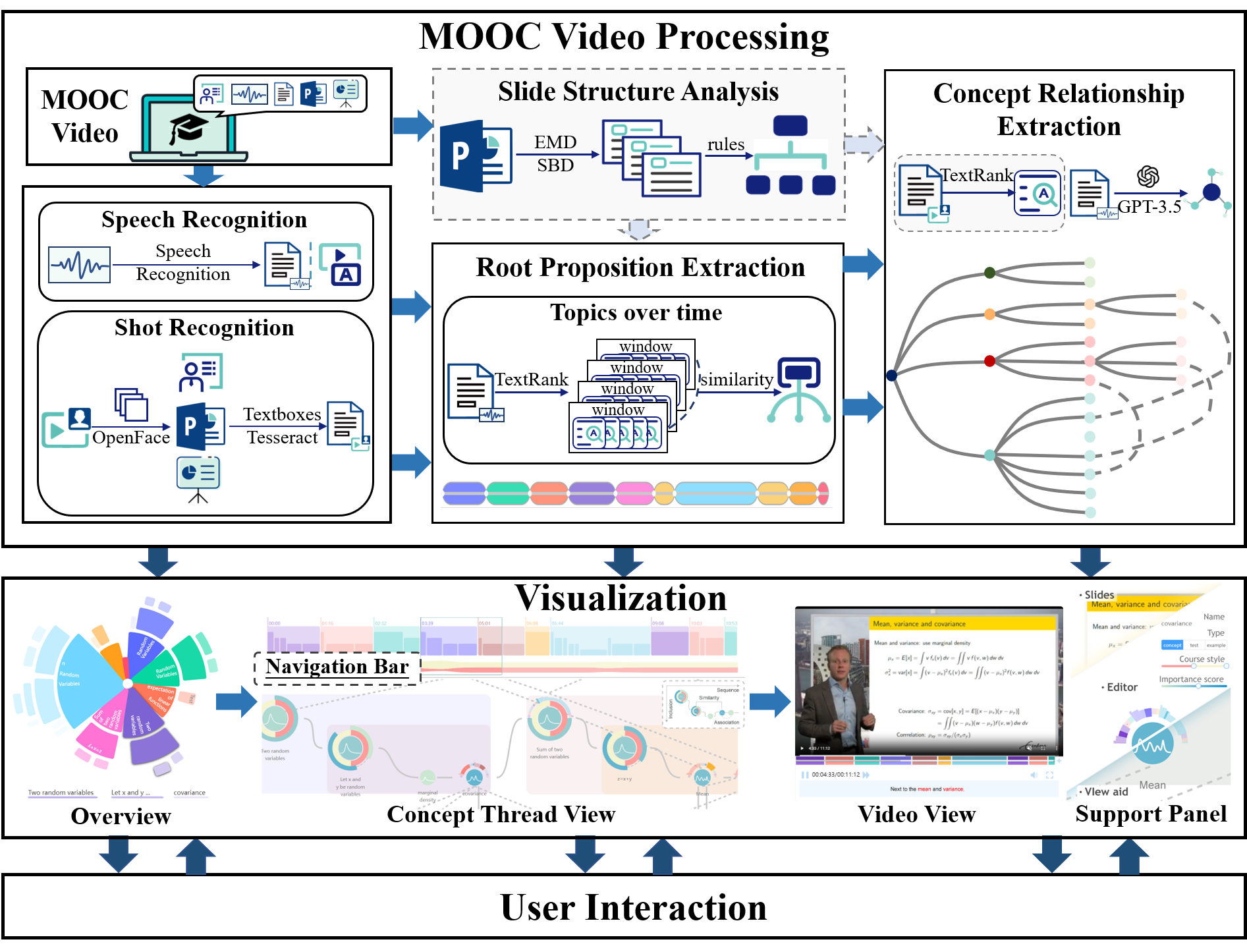}

	\caption{\revise{The system overview of \toolName{} includes a MOOC video processing with four components to extract data and five coordinated views to facilitate interactive exploration of MOOC videos.}}

	\label{Fig3}
\end{figure}

\revise{To fulfill the design requirements mentioned above, we present \toolName{},
a novel visual analytics system
that automatically constructs concept maps from MOOC videos and further visualizes them. 
The system pipeline (Fig.~\ref{Fig3}) consists of four modules: \textit{Speech and Shot Recognition}~\cite{SpeechRecognition,smith2009adapting,liao2017textboxes,baltrusaitis2018openface,chorianopoulos2013usability}, \textit{Root Propositions Extraction}~\cite{mihalcea2004textrank,wang2006topics}, \textit{Concept Relationship Extraction}~\cite{mikolov2013distributed,wei2022chain} and \textit{Slide Structure Analysis}~\cite{seaton2014does,rubner2000earth,zhao2017novel}. 
The \textit{Speech and Shot Recognition} module
uses automatic speech recognition \cite{SpeechRecognition} to convert audio to text and classify shot categories using a recognition algorithm~\cite{baltrusaitis2018openface}. }

\revise{Next, the \textit{Root Proposition Extraction} module employs topic extraction models~\cite{wang2006topics} for extracting root propositions from the input data.
Then, the \textit{Concept Relationship Extraction} module utilizes GPT3.5 and a series of prompts~\cite{wei2022chain} to further extract the concepts and relationships in each root proposition to construct the final concept map.
During this process, the \textit{Slide Structure Analysis} module leverages rule-based headline and slide structure extraction methods to assist in generating results for \textit{Root Proposition Extraction} and \textit{Concept Relationship Extraction}.}

\revise{Based on the multi-coordinated view design principles proposed by Baldonado et al.~\cite{wang2000guidelines}, we designed and implemented \toolName{} interface with five coordinated views to facilitate the exploration of MOOC videos at different learning stages.
The system interface of \toolName{} includes: a \textit{control panel} to display detailed course information and support system configurations; an \textit{overview} to intuitively present the course structure, allowing users to quickly understand the course; a \textit{concept thread view} to display concept and their relationships in detail through a thread metaphor, where the novel glyph design of the concepts explains their importance as well as the temporal distribution and repetitive changes; a \textit{video view} to play the course video to help users quickly navigate the original representation of the corresponding concepts; and a \textit{support panel} displays slides and assists with course exploration. 
To help learners refine the results of the system pipeline, in the edit mode, users can interactively edit concepts and relationships, with changes applied to all views.}

\section{\revise{MOOC Video Processing}}
\subsection{Speech and Shot Recognition}
\label{sec:speech_shot_recognition}
\revise{The teachers' spoken content, \revise{conveyed through} the audio channel, \revise{plays a crucial role in helping students comprehend concepts and relationships.} It serves as a vital source for identifying root propositions.} 
For videos without subtitles, we employ Google's Speech Recognition~\cite{SpeechRecognition} \revise{to transcribe video speeches to text. For videos with subtitles, we directly use the text from them.}

We also employ the visual channel as another primary source for \revise{extracting course content}.
We classify the course style into three aspects \revise{based on previous work}~\cite{chorianopoulos2013usability}, including slides, talking-head \revise{(teacher's head or body appears in the video to make the learner more engaging)}, and drawing-board (teacher annotates in the video allows the learner to better follow the course).
To detect whether the teacher is present (talking-head), OpenFace2.0~\cite{baltrusaitis2018openface} is used to identify facial landmarks in the shot.
To recognize the text in each video frame for slide or drawing-board, we employed Textboxes~\cite{liao2017textboxes}.
Based on the extracted text, we utilized Tesseract~\cite{smith2009adapting} to judge \revise{whether} the extracted text information is handwritten or printed text, which is used to differentiate between slides and drawing-board.

\subsection{Root Proposition Extraction}
\label{sec:root_proposition_extraction}
% After completing data processing by obtaining text from both audio and visual channels, we move on to building the concept map. Initially, our plan was to extract concepts directly from the text. However, we encountered challenges as the concepts were scattered too finely throughout the text, ignoring the teacher's course arrangement and overlooking the important linear expression process. To address this, we use topic modeling to extract the root proposition from the video, providing an outline of the course structure, which we then analyze to establish the relationships between concepts in each section.
After completing data processing by obtaining the text from both audio and visual channels, we use topic modeling to extract the root proposition from the video, which provides an outline of the course structure.

% To extract root propositions, we used the Topics over time~\cite{wang2006topics} model, based on word co-occurrence and time positioning. This algorithm handles timestamp metadata and time segmentation since lecture content is divided into time segments. Firstly, we extracted key phrases and keywords using Textrank~\cite{mihalcea2004textrank} and divided them into different time windows according to the timestamp. Each window contained key phrases recognized in several sentences, with a window size of 5 seconds. For each phrase in each window, we called it a topic and calculated its weight in the adjacent windows before and after. We matched topics between windows using "cross-time analysis" and calculated similarity using cosine similarity. The weights were used to determine the evolution trend of each topic, and after fitting similar topics, we determined the boundaries of root propositions based on weight (Equation 2, where w\_i represents the weight of the topic in time window i, and k is the size of the time window) change over time. The topic with the highest weight in each time period was selected as the root proposition of that period.
% To extract root propositions, we used the Topics over time model~\cite{wang2006topics} based on word co-occurrence and time positioning, since the model is suitable for handling timestamp metadata and time segmentation.
\revise{Given that texts often have timestamp information, we use the TOT (topics over time) model~\cite{wang2006topics} to extract root propositions} based on word co-occurrence and time positioning. 
Firstly, we extracted key phrases and keywords using Textrank~\cite{mihalcea2004textrank} and divided them into different time windows according to the timestamp. 
\revise{The default} window size is 5 seconds.
For each phrase or keyword in each window, we called it a topic and calculated its weight in the adjacent windows. 
In Window $i$ and Window $i+1$, we matched topics between windows using ``cross-time analysis" and calculated their cosine similarity.
The similarity is regarded as the weight of Topic $t$ in Window $i+1$, and used to determine the evolution trend of each topic, and after fitting similar topics, we determined the boundaries of root propositions based on weight change over time. 
The topic with the highest weight in each time period was regarded as the root proposition.

\subsection{Concept Relationship Extraction}
\label{sec:concept_relationship_extraction}
After extracting the root propositions, we obtain the segmented structure of the course content, and then we further extract concepts and relationships \revise{within each root proposition.}

\revise{We utilize the powerful GPT-3.5 and create prompts following the Chain-of-Thought model proposed by Wei et al.\cite{wei2022chain}, we constructed a series of prompts to guide GPT in generating inference-based results step by step and require explicit presentation in the results. First, we characterize the model roles through natural language, such as ``I would like you to play the role of a professional data analyst who excels at constructing concept maps from text". Then, we inputted the text from root propositions and provided prior knowledge by incorporating the keywords extracted from the video. We instructed the model to generate the final results based on the hierarchical structure and return the generated results in a predefined JSON format. Throughout the process, when slides are present in the video, we incorporate the concept map structure obtained through slide analysis as prior knowledge into our prompts and emphasize its consideration. We place the complete prompts in the Appendix for reference.}

We divide the relationships into four categories in the Sec~\ref{sec:task_analysis}:~\textit{Sequence},~\textit{Association},~\textit{Similarity}, and~\textit{Inclusion}. 
The concept map construction enables \textit{inclusion} relationships to be effectively extracted, while ~\textit{Sequence} relationships are generated based on timestamps. 
\textit{Similarity} relationships are identified by similar word embedding vectors, which are extracted based on word2vec~\cite{mikolov2013distributed}.
We define \textit{Association} relationships by searching for corresponding concepts in the audio channel.
\revise{If the teacher mentions other concepts during the time in which he/she is explaining a concept, we establish a \textit{Association} relationship between these two concepts.}

\subsection{Slide Structure Analysis}
In MOOC videos, audio data is the primary information source due to its comprehensive nature. However, visual aids such as slides and drawing-board contain crucial information for accurate concept map generation. To improve the extraction accuracy, we use a \revise{\textit{two-stage SBD strategy} based on Zhao et al.~\cite{zhao2017novel}} to segment slides into shot units, measuring the dissimilarity between frames using Earth Mover's Distance (EMD)\cite{rubner2000earth}, and refining sliding boundaries with the edge-based SBD method\cite{seaton2014does} to identify each slide. Slides are mainly useful in the root proposition and concept relationship extraction stages.

\textit{\uline{Root Proposition Extraction.}}
\revise{To enhance the results in root proposition extraction}, we segment slide topics using \revise{the \textit{rule-based headline extraction} method} by linguistic clues. We measure the textual similarity of slide titles since slides within the same topic usually share common terms. We tokenize the slide titles, and remove stopwords and stems to obtain two word sets. The similarity between adjacent slide titles \revise{is then calculated by measuring the word overlap between these sets.}
Slide titles are extracted using rule-based methods as follows:

\begin{itemize}
	\setlength{\itemsep}{0pt}
	\setlength{\parsep}{0pt}
	\setlength{\parskip}{0pt}
  \item \textit{Location.} The top and left portions of a slide are more likely to contain titles compared to the bottom and right.

  \item \textit{Font size.} Titles usually feature larger font sizes. If all paragraphs have the same font size, we select the top paragraph as the slide's title. 

  \item \textit{Color.} Titles should have the same color throughout all slides. Therefore, a paragraph is selected as a headline only if it has the same color as other titles on all slides.

  \item \textit{Word Count.} The number of words in a title is limited to 1-12. It should be noted that non-alphanumeric strings like numbers and punctuation are not counted.   

  \item \textit{Isolation.} Titles are usually far away from their body text. We calculate the average distance between paragraphs by analyzing the spacing between them. A title should be positioned farther away from its nearest body paragraph than the distance between paragraphs.

\end{itemize}

\textit{\uline{Concept Relationship Extraction.}} 
To refine the \revise{final concept map in Sec.~\ref{sec:concept_relationship_extraction}, we employed a \textit{rule-based slide structure extraction} algorithm to analyze the slide objects}, including text blocks, embedded images, charts, and graphics. An organized hierarchy can be recognized by indentation and font characteristics. To extract structural features of text, we use a bottom-up algorithm to group homogeneous text regions into logical units. Potential bounding boxes of words are detected using TextBoxes~\cite{liao2017textboxes}.

\revise{Then, we group bounding boxes into text lines based on several criteria: same y-coordinate, consistent font size among words, matching character spacing, and a horizontal distance between adjacent bounding boxes not exceeding twice the average character width.
Next, to group text lines into paragraphs, we consider similar bounding box heights and the alignment of adjacent text lines (left, center, or right). We analyze horizontal indentation relationships to extract the hierarchical structure and identify concept phrases. These phrases are found by detecting attributes such as bold, italicized, or underlined text, variations in font size or color, or proximity to punctuation marks (e.g. ``:,.").}

\revise{The hierarchies generated through the slides are used as important prior knowledge to assist in the \textit{Concept Relationship Extraction} stage, thus filling in or correcting relationships or concepts that are not recognized through the script.}

\section{Visual Design}
In this section, we introduce the visual encoding of each view in detail and discuss alternative designs. Then, we describe the user interactions supported in \toolName{}.

\vspace{-2mm}
\subsection{Overview}
The overview~\minorRevision{(Fig.~\ref{fig:teaser}-c)} of \toolName{} is the main entry to explore the video content and 
\minorRevision{provides a macro-level visual summary of the knowledge, where both the knowledge structure and detailed content in the MOOC video are clearly shown (R4).}
% presented in the MOOC video (R5).
% \minorRevision{The visual design aims to succinctly depict the course's structure and content (R4).}
% The visual design aims to provide a clear and concise overview of the course structure and content (R4). 
% We present the course concepts in a sunburst chart that exhibits hierarchical relationships from the center outwards. 
\minorRevision{We utilize a sunburst chart to display the video's hierarchical concepts, expanding outwardly.}
% \wy{Pls check my Chinese comments.}
\revise{To maintain the visual consistency} of root propositions, we utilize a unified color scheme. 
% For quick comprehension,
\minorRevision{Specifically,
the concepts are arranged clockwise in chronological order, and concepts with the same root proposition have a uniform color scheme.}
% the concepts are color-coded in a clockwise manner based on their order of occurrence.
% \wy{I do not understand this sentence. What do you mean by saying ``color-coded in a clockwise manner based on their order of occurence''?}
% To represent time duration and importance, we encode the shape and color \revise{of each sector.
% We utilize shape to represent the duration of concepts, where concepts with longer durations are encoded with larger angles of the sector.}
% For importance, we employ color to represent word frequency, and the outer radius indicates the number of relationships, with darker colors and higher sector height signifying more important concepts. 
\minorRevision{Concept duration and importance are encoded by sector shape and color, respectively.}
% concepts with longer duration have wider sectors,
We utilize shape to represent the duration of concepts, where concepts with longer duration are encoded with larger angles of the sector.
For importance, we employ color to represent word frequency, and the outer radius indicates the number of relationships, with darker colors and higher sector height signifying more important concepts.
% while word frequency and relationship amount are indicated by darker colors and taller sectors. 
% \wy{very confusing here. What are ``longer concepts''? What is the relation between ``concept importance'' and ``frequency and relationship amount''?}
\minorRevision{We limit the maximum height of each hierarchical node to better express the hierarchical structure, facilitating a clear structural understanding and enhancing users' comprehension of the course content.}

% To emphasize higher-level nodes, we use varying height representations since mapping them at the same height would make it challenging to distinguish between levels, and higher-level nodes better illustrate the course structure. 
% This enhances visual performance and allows users to better comprehend the course content.

% a sunburst chart displays the course's hierarchical concepts, expanding outwardly. We ensure visual uniformity of the core ideas through a standard color scheme. For rapid understanding, we arrange concepts clockwise by their sequence, coloring them accordingly. 

\vspace{-2mm}
\subsection{Concept Thread View}
\begin{figure}[tb]
	\centering
    \hspace{-2mm}

	\includegraphics[width=\linewidth]{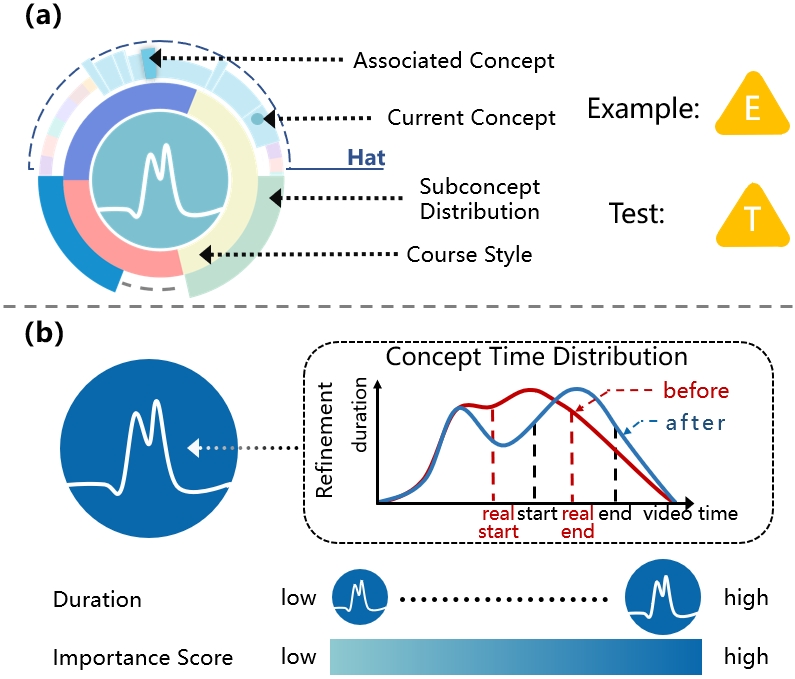}

	\caption{\minorRevision{Visual design for concepts and propositions. (a) shows the radial design for Concepts and Propositions with the associated concept, sub-concept, and course style, along with supplementary design utilized in examples and tests; (b) demonstrates the concept presents three key attributes: duration, importance, and time distribution.}}

	\label{Fig4}
    \vspace{-3mm}
\end{figure}

\begin{figure}[t]
	\centering

	\includegraphics[width=\linewidth]{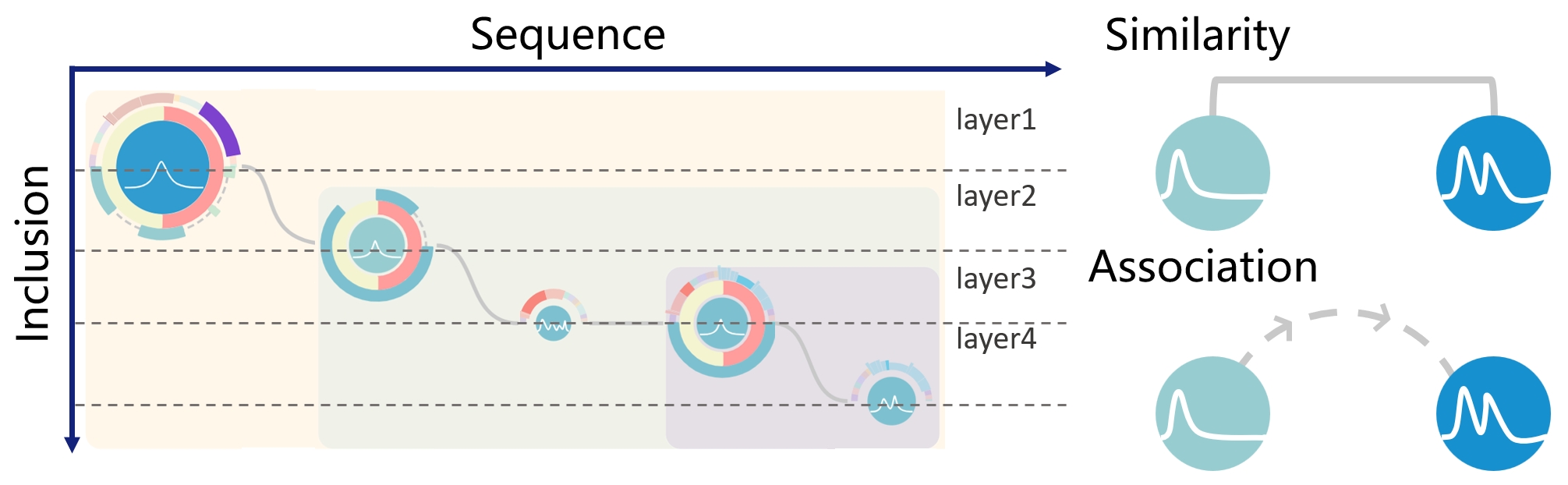}

	\caption{The visual encoding of the four basic relationships.}
	\label{Fig6}
    \vspace{-2mm}
\end{figure}

\revise{\minorRevision{As Nesbit et al.~\cite{nesbit2006learning} summarized}, concept maps are composed of nodes and edges.
Therefore,} \toolName{} offers a concept thread view (Fig.~\ref{fig:teaser}-b) to analyze the threaded concepts through two parts: node design~\minorRevision{(Fig.~\ref{Fig4})} for concepts and propositions and edge design~\minorRevision{(Fig.~\ref{Fig6})} for relationships (R5).

\revise{To help users navigate through the concept map~\minorRevision{(Fig.~\ref{fig:teaser}-b2)}, we create a navigation bar~\minorRevision{(Fig.~\ref{fig:teaser}-b1)} above it, which tightly couples the concept map and the timeline. \minorRevision{The navigation bar incorporates a bar chart and a thematic river. The bar chart encodes linearly, with the heights of the bars indicating different levels of concepts and the same color denoting identical root propositions.
Below the bar chart, a thematic river outlines the course style (R4), segmenting by duration and course style transitions.
% \wy{pls check my Chinese comments.}
This explicit depiction of course style helps analyze teaching method shifts, prompting learners to redistribute attention.
}}

\textbf{\revise{Radial Glyph Design for Concepts and Propositions.}}
To fully illustrate the elements of the concept map, we have categorized the nodes into two types:~\minorRevision{1) concepts and propositions, and 2) knowledge examples and tests, as shown in Fig. \ref{Fig4}-a}.
% The first type represents concepts and propositions, which is the critical part of the concept map. 
\minorRevision{Concepts and propositions are the critical part of the concept map and we propose a radial glyph design to visualize them, as shown on the left side of Fig. \ref{Fig4}-a.
Knowledge examples and tests refer to the examples to illustrate the concept and quizzes or other tests to examine how well students master the knowledge, which are indicated by the uniform yellow triangles labeled with ``E'' and ``T'' respectively, as shown on the right side of Fig. \ref{Fig4}-a.
% The second type represents examples and tests, indicated by uniform yellow triangles labeled with ``E'' and ``T'' respectively. 
% By portraying tests and examples as nodes, users can easily discern how to practically apply the concept they are currently learning.
% This is particularly beneficial for users who prefer to learn concepts through practical examples or tests.
% This visual distinction allows users to easily understand the application of learned concepts, especially benefiting those who favor learning through practical examples or tests.
Such visual designs allow learners to easily navigate to the concepts and propositions of their interest as well as relevant examples or tests, facilitating effective learning. 
}

% \minorRevision{The concepts and propositions are visualized as intuitive radial-based glyphs (Fig. \ref{Fig4}-a) to reduce cognitive load during learning.}
%
%
% To represent concepts or propositions, we have designed radial-based visualizations (Fig. \ref{Fig4}-a). 
% As one of our analysis tasks is to provide course style analysis (R4), if the concepts with sub-concepts, we have an outer ring to the original graphics, which maps the three preset expression modes (slides, talking-head, drawing-board) to the ring based on the duration of their expression within the concept. 
\minorRevision{
Specifically, the radial glyph design (Fig. \ref{Fig4}-a) aims to reduce the cognitive load when learning concepts and propositions.
We design an outer ring for concepts, mapping the duration of three preset expression modes (slides, talking-head, drawing-board) to this ring (R4).
For concepts with sub-concepts, we add an outermost ring with a segmented design, representing the sub-concept distribution,
% \wy{What do you mean here?  representing the sub-concept distribution of \textit{inclusion} relationship concepts????}
%
%
where each color block's size corresponds to the sub-concept's duration, and the color corresponds to the sub-concept's importance score.
% To clarify sub-concept identification, we match block colors to the internal colors of the sub-concepts and sequence them clockwise.
In this way, users can retain detailed information about the sub-concepts, even when the main concepts are collapsed.}

\revise{We add a ``hat'' \minorRevision{(Fig. \ref{Fig4}-a)} to the concepts to indicate their \textit{association} relationships (if any), aiming to reduce visual clutter resulting from an abundance of such associations (R3).
We use a bar chart design, the root propositions to which the concepts are related are fully displayed and explicitly enlarged, with the corresponding concepts highlighted.
\minorRevision{The rest of the root propositions are shown only by themselves and scale down to the smallest recognizable size to provide more visibility.}
A dot within the bar chart serves as an indicator of the concept's current position.}

For the inner part of the radial design, \minorRevision{we show a circle with a sparkline} \minorRevision{(Fig.\ref{Fig4}-b)}. The circle's radius corresponds to the concept's~\minorRevision{temporal length}.
The sparkline displays the time distribution of the concept, indicating its degree of activity (R2).
According to Karpicke et al.~\cite{karpicke2007repeated}, teachers may repeatedly mention important concepts, and obtaining this information explicitly is important for users.
The sparkline design~\minorRevision{illustrates concept occurrence time and duration}, \minorRevision{with the peak height indicating the duration of the concept at different time points}. 
% We do not strictly follow the correct time scale to draw the sparklines to avoid visual clutter.

% \wy{Reach here.}

The circle's background color shows the concept's importance \minorRevision{(Fig.\ref{Fig4}-b)},~\minorRevision{where
% darker shades signifying greater significance 
darker blue signifies greater importance
(R1).}
% We propose an importance score that combines the concept's relationship degree and word frequency, reflecting its activity in the course and interaction frequency with other concepts.
\minorRevision{
% We propose an importance score combining the concept's relationship degree and word frequency, representing its interaction frequency and activity within the course.
The importance score is a weighted combination of the occurrence frequency of the concept and its interaction frequency with other relevant concepts.
The occurrence frequency of different concepts is determined using the TF-IDF algorithm~\cite{salton1988term}, and the interaction frequency with other concepts is determined using the PageRank algorithm~\cite{brin1998anatomy}.
}
% The importance of concepts is obtained by weighting the two components. 
% The degree of the concept relationship reflects its level of connectivity within the concept map, which is calculated using the PageRank~\cite{brin1998anatomy} algorithm.
% The word frequency reflects how frequently the information describing the concept is mentioned in the course, and the score is obtained through the TF-IDF~\cite{salton1988term} algorithm.
% \minorRevision{
% % The score is calculated by weighting these two factors.
% The relationship degree, reflecting concept connectivity in the map, is determined using the PageRank algorithm~\cite{brin1998anatomy}. 
% Word frequency, indicating the concept's mentioned frequency in the course, is assessed via the TF-IDF algorithm~\cite{salton1988term}.}

\textbf{\revise{Concept Relationships Representation.}}
% We aim to minimize the cognitive load on users by linearly organizing the relationships between concepts logically (R3). 
\minorRevision{
% As shown in Fig.~\ref{Fig6}, we strive to reduce users' cognitive load by organizing concept relationships linearly (R3).
Fig.~\ref{Fig6} illustrates how \toolName{} visualizes the relationships between concepts, where different concepts are linked linearly to show their sequence order and inclusion relationship, reducing learners' cognitive load (R3).
}
\revise{Specifically, we \minorRevision{utilize} a thread-based metaphor to convey the \textit{sequence} relationship, which is the most critical design in our approach.}
\revise{We flatten concepts in chronological order and connect them linearly, encoding them horizontally based on time. This serves as the foundation for all other relationships.}
% For the design of the other relationships, as shown in Fig.~\ref{Fig6}.
\minorRevision{For \textit{inclusion} relationships, we visually represent them hierarchically, with the root proposition at the top level and subsequent concepts displayed below.}
% By utilizing horizontal and vertical encoding,
% we offer a visual design with multiple trees with a temporal \revise{feature}. \wy{I do not understand it.}
%
\minorRevision{We enhance the visual representation of \revise{\textit{inclusion}} relationships by adding background color, where all the descendant concepts of a concept are covered by the same background color (Fig.~\ref{Fig6}).}
The \textit{association} \minorRevision{relationships} refer to the mutual deduction or auxiliary explanation between concepts, \minorRevision{which are shown as dashed curves}.
\revise{The \textit{similarity} \minorRevision{relationships} indicate that several concepts share similar results or purposes, \revise{which 
% indicates a strong relationship}, we represent it with \revise{a 
are represented as
solid orthogonal lines}.}

\begin{figure}[tb]
	\centering

	\includegraphics[width=3.5in]{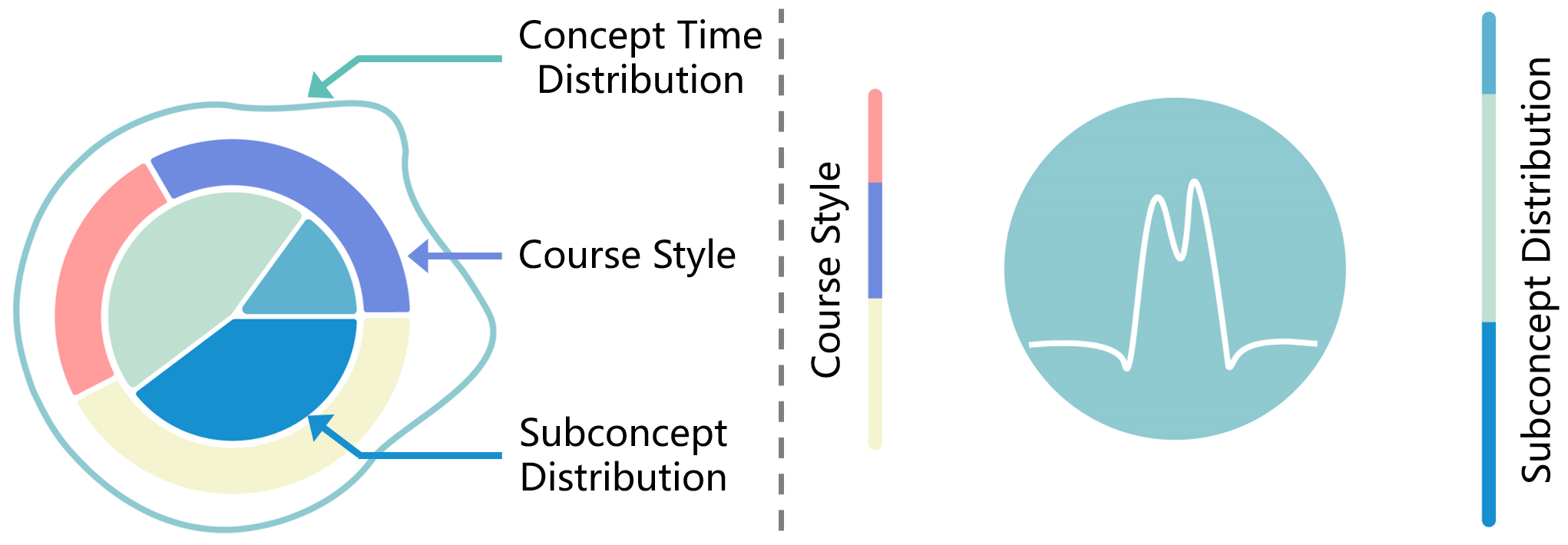}

	\caption{Two alternative designs of Concepts and Propositions}
	\label{Fig5}
    \vspace{-2mm}    
\end{figure}

\textbf{Design Alternatives.} 
During the design process, we carefully consider two alternative glyph designs as shown in Fig.~\ref{Fig5}. \minorRevision{The first alternative design on the left side is to use the pie chart inside to show the sub-concept distribution and the ring around the pie chart to show the concept's time distribution.}
% \wy{Difficult to understand it. Please check my Chinese comments and revise it.}
%
However, this design proved inadequate as the sub-nodes are difficult to see when there are a large number of concepts.
% the concept is too small.
\minorRevision{The ring's angle-based time encoding also hindered intuitive understanding,
and the absence of a unified y-axis 
% complicated duration comparisons.
makes it hard to compare the durations of different concepts.}
The second alternative design on the right side encodes the sub-concept distribution and the course style by the lengths of two vertical bars.
% \wy{which part of which figure? The right part of Figure 5???}
However, this design may become difficult to discern when the vertical bars are too short, \minorRevision{and can lead to more discrete nodes, impacting the threaded-based design.}

\vspace{-2mm}
\subsection{Interactions}

\toolName{} enables a set of interactions to help users explore MOOC videos through multiple views.

\textbf{\revise{Linking.}}
The system maintains cognitive consistency by linking all views.
when a concept is selected, \minorRevision{users} can click on it to get video navigation, all views are synchronized to show the current concept, with the support panel providing the corresponding slides. 
Alternatively, when the video is played, the concept thread view moves horizontally to locate the current concept.
The internal sparkline of concepts provides a new navigation method, with a red dot appearing when approaching a peak, supporting jumps to similar concepts.

\textbf{\revise{Demonstrating.}}
Based on the knowledge cognitive model~\cite{AboutLearning}, \toolName{} provides a series of interactions to demonstrate different stages of knowledge acquisition.
When a concept is selected, \revise{all the relationships are displayed, and the navigation bar will also synchronize and highlight the associated concepts,} helping users link the newly-acquired knowledge with \minorRevision{their existing knowledge or concepts.}
For relationships, the support panel displays the concepts \minorRevision{that are currently under review} for a more detailed exploration.
Finally, users can 
% apply their existing knowledge by 
further enhance their understanding of a specific concept by
navigating to examples and tests in the same course video, or referring to other relevant concepts.

\textbf{\revise{Refinement.}}
We've introduced an interactive edit mode (R6).
Users can correct root proposition errors by adding new ones using the progress bar (Fig.~\ref{fig:teaser}-Progress Bar).
They can also adjust proposition lengths dynamically for modifications or deletions. 
After altering the root proposition, users can further edit concepts by clicking on them, utilizing the support panel as an editing panel. Concepts can be deleted or added through right-click actions.
In real-life scenarios, administrators can review and share user-modified concept maps, enabling ongoing improvements with minimal effort, thus compensating for data processing limitations.

\subsection{\revise{Example Use Scenario}}
As shown in Fig.~\ref{fig:teaser}, consider Alex, \minorRevision{a student prepping for final exams. 
% with only two weeks left.
}
\minorRevision{With} \toolName{}, \minorRevision{he can quickly get the concept outline of the MOOC video from the} \textit{overview} (Fig.~\ref{fig:teaser}-c).
% showed an outline of the video, which he quickly skims to get a rough concept awareness. 
Then, he shifts his attention to the \textit{concept thread view} (Fig.~\ref{fig:teaser}-b) \minorRevision{that displays the detailed \minorRevision{concept} information and their interrelationships.}
During the exploration, \minorRevision{he realizes that he doesn't need to watch the entire video to understand the underlying content.}
He \minorRevision{focuses on} the dark-colored concept of ``covariance'', which is implied to be important, and further dives into the \textit{video view} (Fig.~\ref{fig:teaser}-d) \minorRevision{for a detailed study}.
However, he finds it challenging to grasp this concept directly. Through interaction with the concept, he discovers that ``mean'' and ``variance'' are prerequisites for understanding ``covariance'', and ``covariance'' is examined in the subsequent ``tests''.
\minorRevision{
% This knowledge cognitive path 
Such relationships between the concept ``covariance'' and other concepts
are essential for comprehending ``covariance'', with all related concepts synchronized in the \textit{support panel} (Fig.~\ref{fig:teaser}-e).}
\minorRevision{Finally, he employs the sparkline to review ``covariance'' occurrences} to
get a sense of its importance in the whole MOOC video.
% better retain the concept in working memory, aiding in the retention of long-term memory.

% Struggling with “covariance,” he learns through interaction that “mean” and “variance” are essential precursors, and “covariance” is vital in subsequent “tests”. 
% This knowledge cognitive path is essential for comprehending “covariance,” with all related concepts synchronized in the support panel (Fig. 1-e). 
% Finally, he employs the sparkline to review ``covariance'' occurrences for better retention in working and long-term memory.

\section{Evaluation}
Our system is a web application implemented using Flask, VueJS, and D3.js.
To evaluate its usefulness and effectiveness, we conducted a quantitative study, a case study, and a user study using real-world MOOC videos.

\subsection{\revise{Quantitative Study}}
\revise{We conducted a series of experiments to evaluate the three components of data processing: \textit{Root Proposition Extraction}, \textit{Concept Relationship Extraction}, and \textit{Slide Structure Analysis}.
The experimental videos are from \textit{Coursera}\footnote{\url{https://www.coursera.org}}.
We randomly sampled 50 videos (25 with slides and 25 without slides) and these 50 videos are used in our experiments for evaluation.}
\minorRevision{
Three coauthors of this paper worked together to manually construct the ground-truth labels for root proposition, concept relationships and slide structures, which was achieved via reviewing the original videos, scripts, and course descriptions.
% The ground truth for the video was collaboratively developed by three authors in the author list. By reviewing the original videos, scripts, and course descriptions, we established ground truth for key data processing components.
For all videos, two coauthors manually extracted the list of root propositions (including name and time). For videos with slides, the two coauthors further extracted three components: the final state of each slide, the list of slide headlines (including name and time), and the slide structure (the tree structure under each root proposition).
Then, the third coauthor will compare the labels by the prior two coauthors and used the common labels as the ground-truth labels. For the cases with inconsistent labels from the two prior coauthors, all three authors further discussed them one by one and finalized the final labels, ensuring the reliability of the final ground-truth labels.
% The final ground truth was derived by comparing and refining the results of each author.
%
%
% Subsequently, we validated the accuracy of our constructed ground truths by inviting two experts (E1 \& E2) from our formative study for a review, ensuring their reliability for the final quantitative study.
% \wy{1. Are you sure you have done this? 2. It does not provide enough details on the validation by the two experts.}
%
%
% \wy{Pls check if my above revisions changed your original idea.}
}

\revise{Due to the presence of various concepts and interrelated relationships within the course, we employed Precision, Recall, and F1-score\minorRevision{~\cite{van1979information}} as metrics in our experiments. Precision assesses the consistency between the model-generated results and the ground truth within the course. A high precision indicates that the model can express course content and relationships more accurately. Recall can evaluate the model's adaptability. A high Recall score means that the model provides results that better cover the course content.}
\minorRevision{F1-Score is a single metric that considers both precision and recall. A high F1-Score indicates that the model is both accurate and thorough in capturing the essential content of the course.}

\textit{\uline{\revise{Slide Structure Analysis.}}}
\revise{
% This stage aided in the generation of the other two stages, 
% \wy{What do you mean here?}
% we first evaluated this stage as shown in Table~\ref{table:Slide Structure Analysis}.
We first assessed the performance of slide structure analysis (Table~\ref{table:Slide Structure Analysis}).
Specifically, the two-stage SBD algorithm~\cite{seaton2014does} achieves an F1-score of 0.909, when considering distinctive shot types in MOOC videos, thereby demonstrating our approach's ability to distinguish different slides.
% for subsequent analysis.
Rule-based Headline Extraction (F1-score: 0.891) outperformed Rule-based Slide Structure Extraction (F1-score: 0.872), because it is more difficult to identify concept relationships than to extract root propositions using slide titles.
}
% but the final results demonstrate that the model has effectively struck a balance between precision and recall, successfully extracting the titles within the slides and further analyzing their structure.}
\minorRevision{At this stage, a higher Precision means the model
% is more accurate in identifying slide content, 
can identify the key slide content more accurately, reducing the necessity for learners to revisit the original video to check the actual key slide content.
% thereby better guiding the correct creation of course content during the \textit{Concept Relationship Extraction} stage. This allows learners to spend less time revisiting the original video to correct misunderstandings. 
% Whereas lower Precision may include irrelevant or incorrect slide structures, leading to confusion or misconceptions about the course content.
% For Recall, 
A higher Recall can ensure that the slide structure is more adequate and complete, 
% aiding in producing results closer to the original course, thus better demonstrating the system features, 
helping learners more completely grasp the major concepts in a course video.
% to achieve better learning outcomes, and enhancing their grasp of the overall course. A decrease in Recall may miss crucial structural elements, leading to an incomplete understanding of the course content.
}
% \wy{Pls check my Chinese comments.}

\begin{table}[t]
\centering
\caption{\revise{The performance of \textit{Slide Structure Analysis} stage.}}
\label{table:Slide Structure Analysis}
\begin{tabular}{@{}cccc@{}}
\toprule
                                      & Precision & Recall & F1-Score \\ \midrule
Two-stage SBD method                  & 0.932     & 0.893  & 0.909    \\
Rule-based Headline Extraction        & 0.908     & 0.878  & 0.891    \\
Rule-based Slide Structure Extraction & 0.892     & 0.863  & 0.872    \\ \bottomrule
\end{tabular}
\vspace{-4mm}
\end{table}

\begin{table}[t]
\centering
\caption{\revise{The performance of \textit{Root Proposition Extraction} stage.}}
\label{table:Root Proposition Extraction}
\begin{tabular}{@{}cccc@{}}
\toprule
                                      & Precision & Recall & F1-Score \\ \midrule
With Slides                  & 0.878     & 0.901  & 0.868    \\
No Slides (TOT~\cite{wang2006topics} )        & 0.759     & 0.807  & 0.782    \\
\bottomrule
\end{tabular}
\vspace{-4mm}
\end{table}

\textit{\uline{\revise{Root Proposition Extraction.}}}
\revise{We further evaluated the performance of \textit{Root Proposition Extraction} with and without slide assistance (Table~\ref{table:Root Proposition Extraction}). 
Due to the instability of the TOT model's \cite{wang2006topics} in clustering topics in the time series, root proposition extraction without slides achieved an F1-score of 0.782. 
% \wy{pls check my Chinese comments.}
However, by incorporating the content of slides, the overall performance of \textit{Root Proposition Extraction} is significantly improved (Precision Increase: 11.9\%, Recall Increase: 9.4\%, F1-score Increase: 8.6\%). 
%
%
% The information within the slides can greatly assist in compensating for the shortcomings of the TOT model, while demonstrating the effectiveness of the \textit{Root Proposition Extraction}.
}
~\minorRevision{At this stage, compared to Precision, changes in Recall more significantly impact the learning process. Based on user feedback, they prioritize whether the system's root propositions comprehensively represent the course outline, with less emphasis on the correctness of these propositions. Quick retrieval of the original video can compensate for lack of accuracy, but lower Recall might mean critical propositions are overlooked, leading learners to prefer exploring the original video directly, thus neglecting the system's features.}
% \wy{NEED AN ONLINE MEETING: 1) What do you want to say here? What is the purpose? 2) Can we really use precision and recall to do the evaluation here (Why not directly use ``accuracy''?)? How are they calculated here?}

\textit{\uline{\revise{Concept Relationship Extraction.}}}
\revise{Evaluating the results generated by GPT-3.5 is a challenging task rooted in two primary factors. Firstly, the concept maps of Mooc videos currently have no ground truth. Secondly, it is difficult to evaluate generative models without ground truth.
%
%
% \wy{Pls check my Chinese comments.}
%
%
Therefore, we invited experts \minorRevision{(E3 \& E4)} to rate 50 concept maps generated by the system on a 1-to-7 scale (from ``totally unacceptable'' to ``very good'').
% \wy{Pls confirm it.}
\minorRevision{E3 is an educational theory researcher interested in analyzing course organization in online education;
E4 is a professor of educational psychology and is researching course delivery methods in online learning.
The experts explored three MOOC video lessons from \textit{Coursera}.}
The results indicated the effectiveness of our approach in constructing concept maps (No Slides: 5.2(E3), 5.4(E4); With Slides: 5.8(E3), 6.1(E4)).}

%\vspace{-1mm}
\subsection{Case Study}
%\vspace{-1mm}

\revise{In this section, we describe how \toolName{} can be used in practice to explore and analyze MOOC videos through two case studies with two experts
\minorRevision{(E3 \& E4).}
% \wy{What do you mean by saying ``Consistent with the quantitative study''?}
}
\revise{In the first case, E3 explored two lessons of the same course: \textit{Data Structure: Chaining}\footnote{\url{https://www.coursera.org/learn/data-structures/lecture/2yY2h/chaining}} and \textit{Data Structures: Chaining Implementation and Analysis}\footnote{\url{https://www.coursera.org/learn/data-structures/lecture/dWNAc/chaining-implementation-and-analysis}}, where the first lesson is a theoretical lesson and the second is the corresponding practical lesson. He aimed to investigate if \toolName{} can effectively explore different types of courses. In the second case, E4 was presented with two lessons, \textit{Information Visualization: Fundamental Graphs}\footnote{\url{https://www.coursera.org/learn/information-visualization-fundamentals/lecture/On8JK/fundamental-graphs}} and \textit{Introduction to Statistics: Descriptive Statistics and Visualizing Information}\footnote{\url{https://www.coursera.org/learn/stanford-statistics/lecture/k4w4I/pie-chart-bar-graph-and-histograms}}, both covering the same content but taught by different teachers. This allowed an analysis of how \toolName{} influences learners' course selection based on their preferences in the learning process.}

\begin{figure*}[t]
    \centering
    \includegraphics[width=\linewidth]{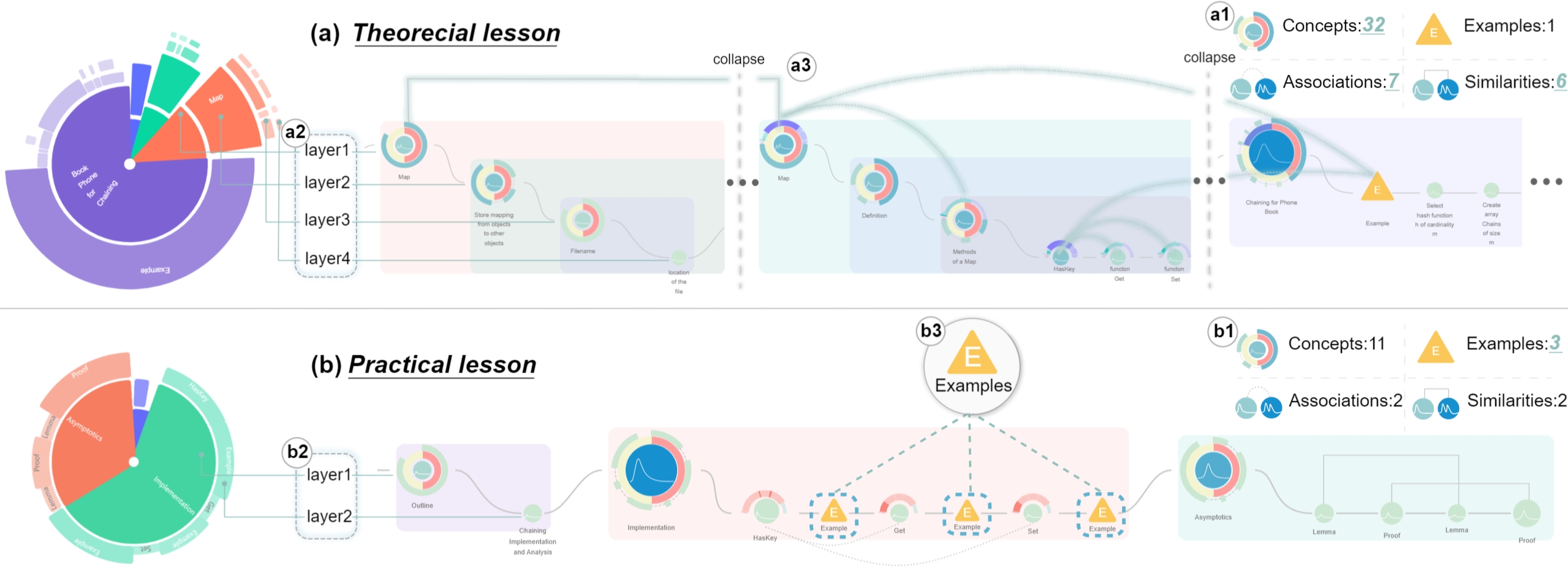}
 
    \caption{\revise{Visualizations of MOOC videos for different course types in the same course. (a) \textit{Data Structure: Chaining} and (b) \textit{Data Structures: Chaining Implementation and Analysis}, where (a) is a theoretical lesson and (b) is the corresponding practical lesson, the \textit{overview} and \textit{concept thread view} are shown in the left and right parts.}}

    \label{FigCase1}
    \vspace{-3mm}
\end{figure*}

\textbf{\revise{Case 1: Different lessons of the same course.}}
\revise{According to Gillies' work on Cooperative learning~\cite{gillies2007cooperative}, combining theory and practice is highly advocated for teaching and learning, thus E3 decided to analyze the characteristics of different types of lessons in terms of their concepts and relationships.}

\textit{\uline{\revise{Concept and proposition exploration.}}}
\revise{E3 initially compared the differences in concepts and propositions between theoretical and practical lessons.
For quantity, the theoretical lesson has more knowledge points (Fig.~\ref{FigCase1}-a1), which are more oriented toward concepts and propositions, and has only a few examples.
the practical lesson, on the other hand, often uses examples to express practical content (Fig.~\ref{FigCase1}-b3), and E3 argued that this is the most obvious difference between the two types of lessons, and that \toolName{}'s distinction between concepts and examples serves this feature well. }

\revise{During further exploration, E3 found that compared to the practical lesson, the theoretical lesson has a deeper hierarchical structure (Fig.~\ref{FigCase1}-a2), which makes the content more abstract and comprehensive, this kind of structure can help beginners understand the video, and also serves well for reviewers to organize their knowledge. 
In contrast, the practical lesson typically features a flatter course structure (Fig.~\ref{FigCase1}-b2), allowing learners to access practical content of interest more quickly.}

\textit{\uline{\revise{Relationship exploration.}}}
\revise{For relationships, E3 found that the theoretical lesson tends to have more \textit{association} relationships to refer to knowledge cognitive paths due to the tight logical structure (Fig.~\ref{FigCase1}-a3), whereas the practical lesson tends to consist of fewer relationships, which was in line with E3's expectations as practical lessons focus on the application of knowledge rather than the illustration of interrelationships. 
For \textit{similarity} relationships, the same concepts are repeated in the theoretical lesson to deepen knowledge, whereas the practical lesson is less characterized by this feature.}

\revise{Through exploration, E3 found that \toolName{} can well distinguish the different patterns of these two course types, and this feature can well help learners at different stages to choose different learning purposes, thus facilitating rapid learning. 
For beginners, \toolName{} provides a complete representation of the concepts, so that learners can complete the theoretical lessons more quickly with the aid of the system, and then go into the practical lessons to apply the knowledge.
For learners reviewing for exams, \toolName{} extracts the structure and logical relationships of the theoretical lessons so that learners can quickly recall what they have learned without having to explore in-depth videos, and provides navigation through the practical lessons to help learners quickly locate the knowledge points they want to review.}

\begin{figure*}[t]
    \centering
    \includegraphics[width=\linewidth]{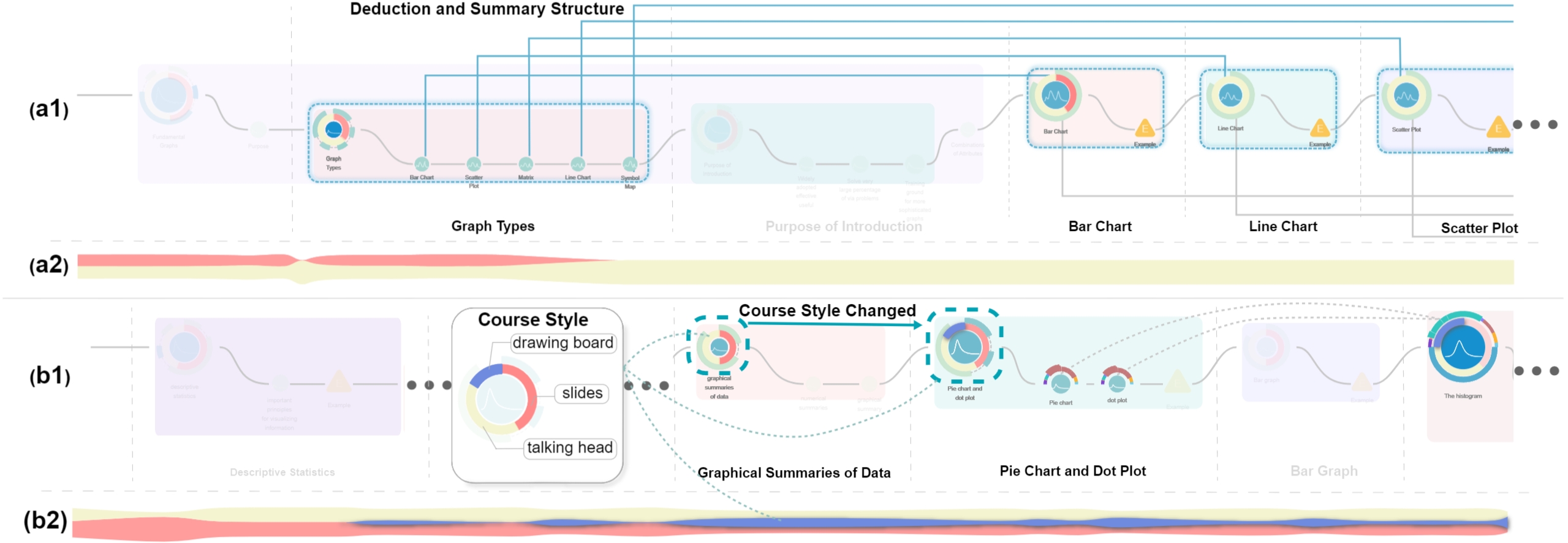}
 
    \caption{\revise{Visualization of MOOC videos with the same content taught by different teachers. (a) \textit{Information Visualization: Fundamental Graphs} and (b) \textit{Introduction to Statistics: Descriptive Statistics and Visualizing Information}, where the respective \textit{concept thread view} and course style theme river are shown in the upper and lower parts.}}

    \label{Fig7}
    \vspace{-2mm}
\end{figure*}

\textbf{\revise{Case 2: Same course taught by different teachers.}}
E4 explored the video on \textit{fundamental graphs} in information visualization. Upon initial exploration of the entire course structure, E4 noted that the teacher introduced various types of basic charts and explained their characteristics in turn, all connected by \textit{similarity} relationships, which means the ``deduction and summary'' structure of the video (Fig.~\ref{Fig7}-a1). E4 noted that \toolName{} helps learners distinguish the course structure, but too many relationships could lead to visual confusion. E4 suggested genetic algorithms as a possible solution for this issue.

\textit{\uline{\revise{Course structure exploration.}}}
\revise{E4 showed great interest in analyzing the diverse teaching approaches used by different teachers for the same content, recognizing the significant impact of instructional variations on learner attention allocation.} She explored \textit{Descriptive Statistics and Visualizing Information} in Introduction to Statistics and compared it with \textit{fundamental graphs}. After initial observations, she discovered that unlike the \textit{fundamental graphs}, the organization of the \textit{Descriptive Statistics and Visualizing Information} tends to be linear. This led to learners needing more working memory for higher-level course expressions.

\textit{\uline{\revise{Course style exploration.}}}
However, upon careful examination of the course style, E4 discovered that in the \textit{fundamental graphs} (Fig.~\ref{Fig7}-a2), the course expression remains stable, completed through the teacher's explanation of slides. \revise{She believed that focusing solely on course style makes it challenging to identify specific time periods requiring focused attention, as the information conveyed throughout the video remains constant. Attention shifts could only be captured by recognizing cues from examples or test prompts.} E4 found this approach inappropriate and highlighted that \textit{Descriptive Statistics and Visualizing Information} addresses this issue effectively. In this video (Fig.~\ref{Fig7}-b2), the teacher adds many board annotations at various times on the basis of oral explanations and slides, which can re-engage learners' attention and many learners enjoy this style of delivery. She pointed out, ``In this video, I can know when the teacher wants me to redistribute my attention, and the learners can anticipate the timing of course style transitions, thereby reminding them to focus their attention.'' She also added that the course style distribution in each concept's outer ring (Fig.~\ref{Fig7}-b1) can be used to identify the specific time of transition.
E4 also mentioned in subsequent interviews that some learners prefer to study courses through examples and tests. \toolName{} explicit presentation of this information is highly beneficial to learners.

\subsection{User Study}

\begin{figure*}[t]
	\centering

	\includegraphics[width=0.84\linewidth]{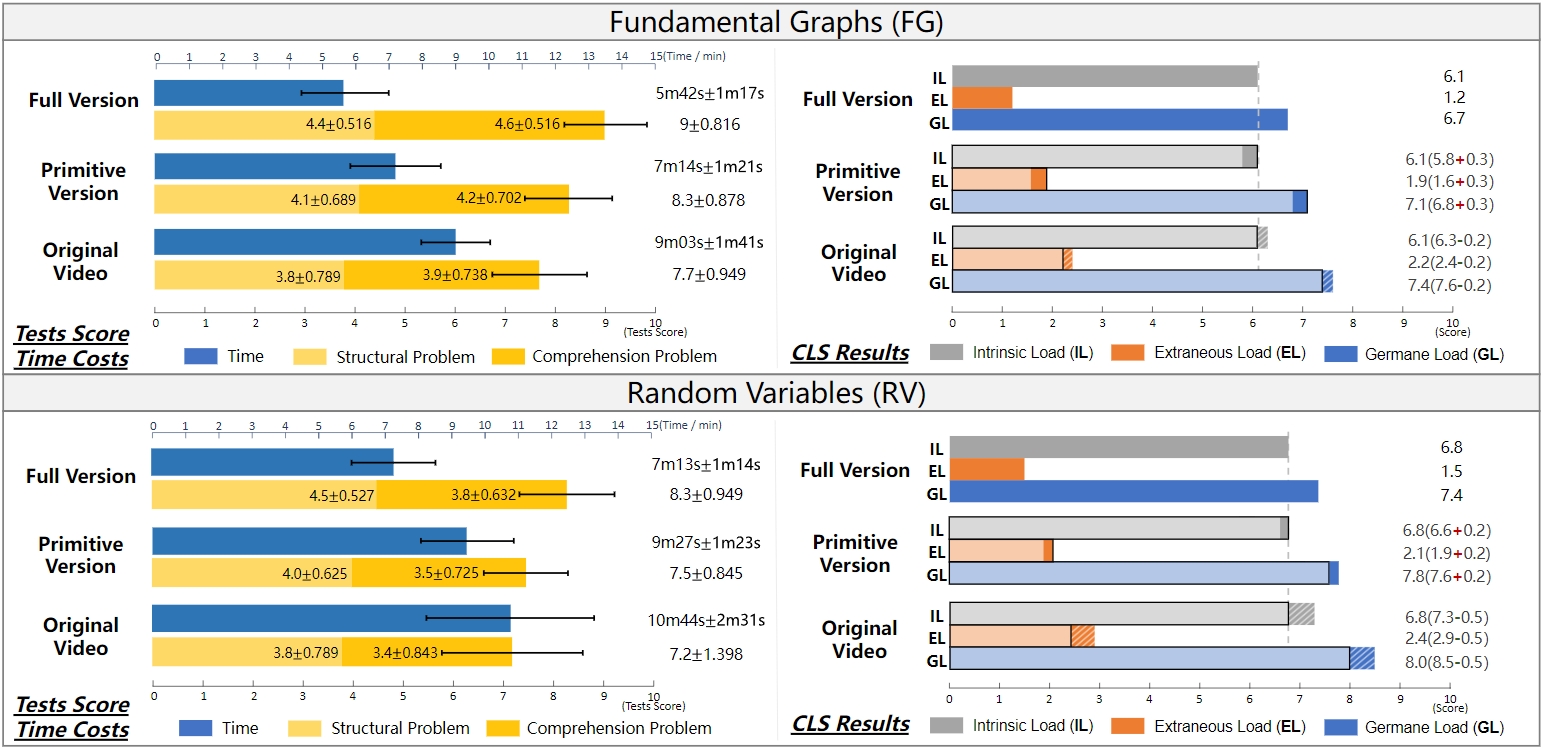}

	\caption{\minorRevision{Results from post-lesson tests (left) and CLS survey (right), the time costs and test scores are located in the upper and lower parts of the bar chart, and the CLS survey showing results after score alignment by intrinsic load (IL).}}
    \label{Fig8}
    \vspace{-3mm}
\end{figure*}

\revise{We further conducted a user study to evaluate the effectiveness and usability of \toolName{}.
Our evaluation focuses primarily on comparing the \toolName{} to our baseline versions using three key metrics: post-lesson tests representing learning outcomes, cognitive load survey~\cite{leppink2013development} reflecting cognitive load during the learning process, and the overall usability assessed via post-study questionnaires.}

\textbf{\revise{Study Design.}}
\revise{To evaluate the \textbf{effectiveness} of \toolName{}, we compared the \toolName{} (namely the \textit{full version}) with a simplified version of \toolName{} (namely the \textit{primitive version}) and the \textit{original video}. There are two differences between the \textit{primitive version} and the \textit{full version}: 
(1) the \textit{full version} provides a complete concept representation, including both \textit{content} and \textit{attributes}, along with four types of relationships, whereas the \textit{primitive version} provides only the concepts themselves and timeline-based \textit{sequence} relationships.
(2) The \textit{full version} uses a range of visual designs and interactions to guide concept cognition, whereas the \textit{primitive version} shows only timeline-based concept entities.}

\revise{We employed a series of methods to assess \toolName{}. Firstly, we conducted post-lesson tests to objectively evaluate participants' learning outcomes. Secondly, we used the Cognitive Load Survey (CLS)~\cite{leppink2013development} to gauge participants' subjective cognitive load. Finally, based on Weibelzahl's evaluation method of adaptive system~\cite{weibelzahl2020evaluation}, we leveraged the four-level taxonomy to create a post-study questionnaire to comprehensively evaluate our system. 
Specifically, we assessed the \textit{informativeness} of the provided video content, the effectiveness of \textit{knowledge acquisition}, the \textit{visual design} of the system, and the overall \textbf{usability} of the system.}

\textbf{\revise{Participants.}}
\revise{We recruited 30 participants (P1-P30, 18 males, 12 females) with a mean age of 22.8 (SD = 2.20) from local universities majoring in computer science, mathematics, art design, education, and finance. Each discipline included two undergraduate and two graduate students. All participants had a basic understanding of MOOCs, and 19 out of 30 had used at least one online learning platform (such as Coursera, edX, and Udacity) for at least 2 years. At the end of the study, we paid \$6 per person for their participation.}

% \wy{Reach here.}

\revise{The participants were divided into three groups.
The first group \minorRevision{(G1)} used \textit{full version} of \toolName{}. The second group \minorRevision{(G2)} studied video lessons using the \textit{primitive version} of \toolName{} that uses a truncated design but with the same concept map structure (the navigation bar shown in Fig.~\ref{fig:teaser}). The last group \minorRevision{(G3)} directly learned from the \textit{original video} without any additional aids.}
\minorRevision{
% Participants from various fields of study may bring prior knowledge to test videos, potentially achieving better test scores. 
% Additionally, their differing experiences with MOOCs could affect their system comprehension speed, course learning time, and feedback on the learning process. 
To ensure balance among groups,
% we stratified participants based on their field of study and MOOC familiarity, then randomly assigned them to one of three groups.
participants were categorized based on their field of study and MOOC familiarity.
They were first classified by their major, then by MOOC familiarity: never used, less than 1 year, and more than 1 year.
% Their field of study matched their major, and MOOC familiarity was classified as never used, less than 1 year, and more than 1 year.
When the results could not be evenly distributed, we replaced some participants to ensure balanced groups, ensuring random allocation into one of three groups.}
% \wy{The new content above does not really clarify how you handle the issue! Pls check my Chinese comments on Overleaf.}

% This stratification aimed to minimize the confounding effects of educational backgrounds and online learning platform experiences.}

\textbf{\revise{Procedures.}}
\minorRevision{
We briefly introduced the research background first, including the research motivation and the study protocol. 
% Next, in addition to the group using the \textit{original video}, to demonstrate how to explore MOOC videos using different versions of \toolName{}, we went through the example of \textit{Case 1} while also describing the visual design, interactions, and workflows. 
Then, we introduced the visual designs and interactions of \toolName{}, and further went through one example on how to use \toolName{} to explore MOOC videos.
After that, the participants were allowed to freely explore \toolName{} to make themselves familiar with \toolName{}. 
% The tutorial above lasted about 15 minutes.
% Each following session lasts approximately 20 minutes and each participant is required to study two lessons in turn with the provided system.
Further, we asked each participant to use either the \textit{full version} or \textit{primitive version} of \toolName{} to explore two MOOC videos, i.e., the videos entitled \textit{fundamental graphs (FG)} and \textit{random variables (RV)} that were used in our above case study.
Among them,
the \textit{fundamental graphs} features a flatter, simpler structure and content (Fig.~\ref{Fig7}-a1), indicating a simple course.
In contrast, the \textit{random variables} encompasses a wider range of concepts and relationships, exhibiting a more complex course structure and content (Fig.~\ref{fig:teaser}), indicating a complex course.
Through research on these two course videos, we can gain a better understanding of how \toolName{} performs in both simple and complex courses.
}

Participants were informed about the lesson names in advance, and only six math majors were familiar with \textit{random variables} lesson. No time limits were set for studying the lesson, but time was recorded and taken into account. We analyzed the learning outcomes when participants indicated that they had finished learning for each lesson.
Participants were asked to answer ten questions, five questions on structural aspects, and five questions on comprehension to objectively evaluate the learning outcomes.
We placed the questions in the Appendix for reference.
Additionally, we conducted a cognitive load survey (CLS)~\cite{leppink2013development} to measure subjective cognitive load during the learning task.
After finishing the post-lesson tests and the CLS, participants proceed to the next lesson. Throughout the study, additional materials (such as paper and the Internet) were prohibited to minimize cognitive load influenced by external tools.
% \wy{pls use consistent tense for all the verbs.}

\revise{After all participants completed studying the two video lessons, we utilized a post-study questionnaire on a 7-point Likert scale (1: not at all, 7: absolutely) from the existing literature~\cite{zhao2017novel,rossi2018evaluation,xia2019peerlens} to gauge user experience and \toolName{}'s usability.
}

\minorRevision{\textbf{Hypothesis.}
We proposed the following hypotheses based on the existing literature~\cite{du2017finding} on peer-based learning.
We then verified the validity of all hypotheses under varying course structure complexity and course content difficulty, specifically for simple courses represented by \textit{fundamental graphs} (\textit{SIM}) and complex courses represented by \textit{random variables} (\textit{COM}).
}

\vspace{-4mm}
\minorRevision{\textbf{\textit{H1.}}
The proposed visual design of \toolName{}, regardless of the full or primitive version, is better than the original video in facilitating effective learning.
Specifically, \toolName{} systems provide higher test scores (\textit{H1a-SIM}, \textit{H1a-COM}) and lower time costs (\textit{H1b-SIM}, \textit{H1b-COM}) compared with the original video.}

\minorRevision{\textbf{\textit{H2.}}
The proposed visual design of \toolName{}, regardless of the full or primitive version, is better than the original video in reducing cognitive load on learning.
Specifically, \toolName{} systems provide less extraneous load (\textit{H2a-SIM}, \textit{H2a-COM}) and germane load (\textit{H2b-SIM}, \textit{H2b-COM}) compared with the original video.}

\minorRevision{\textbf{\textit{H3.}}
The full version of \toolName{} provides more effective learning outcomes than the primitive version. 
In particular, the full version leads to higher test scores (\textit{H3a-SIM}, \textit{H3a-COM}) and lower time costs (\textit{H3b-SIM}, \textit{H3b-COM}) than the primitive version.}

\minorRevision{\textbf{\textit{H4.}}
The full version of \toolName{} is more effective at reducing cognitive load than the primitive version.
In particular, the full version demonstrates lower extraneous load (\textit{H4a-SIM}, \textit{H4a-COM}) and germane load (\textit{H4b-SIM}, \textit{H4b-COM}) than the primitive version.}

\begin{figure}[t]
	\centering

	\includegraphics[width=\linewidth]{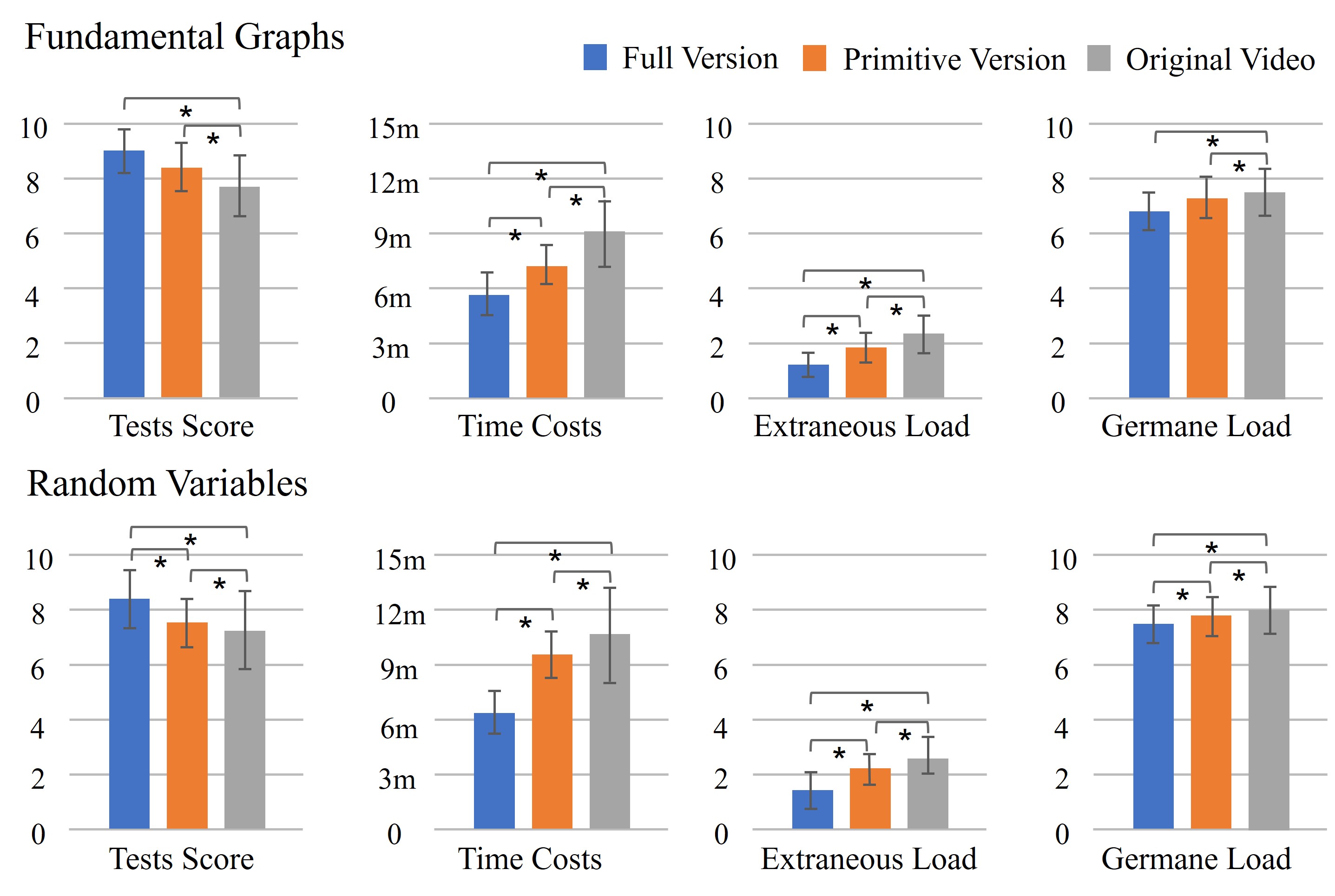}
    \vspace{-6mm}
	\caption{\minorRevision{Means and standard errors of \textit{full version}, \textit{primitive version}, and \textit{original video} on tests score, time costs, extraneous load, and germane load (*: $p<0.05$).}}
    \label{FigSig}
    \vspace{-3mm}
\end{figure}

\textbf{\revise{Results.}}~\revise{We report the participants' quantitative ratings and verbal feedback from three aspects: scores on the post-lesson test, time spent taking the lesson, and results from the cognitive load survey.}
\minorRevision{We run one-way ANOVA on each aspect, followed by the Bonferroni post-hoc test on measures with statistically significant differences.} 
% \wy{Why one-way ANOVA instead of Repeated Measure ANOVA?}

\textit{\uline{\revise{Tests Score.}}}
\revise{Fig.~\ref{Fig8} shows the results of the post-lesson test for each group, with higher scores indicating better learning outcomes.
\textit{G1} (FG: 9.0, RV: 8.3) and \textit{G2} (FG: 8.3, RV: 7.5) achieved higher test scores compared to \textit{G3} (FG: 7.7, RV: 7.2),
\minorRevision{(Fig.~\ref{FigSig}, \textbf{H1a-SIM supported}, \textbf{H1a-COM supported}).}
% \wy{The revised result reported below does follow the general style of statistical difference result report. Pls check my Chinese comments.}
% Table~\ref{table:ANOVA} shows the results of the significance tests, indicating significant differences among different versions (FG: $Sig.=0.01$, RV: $Sig.=0.01$).
% The follow-up Bonferroni post-hoc test revealed significant differences between \textit{G1} and \textit{G2} ($p<0.05$, \textbf{H3a-FG supported}; $p<0.05$, \textbf{H3a-RV supported}) of test scores.
\minorRevision{In the \textit{fundamental graphs} lesson, no significant difference was found between \textit{G1} and \textit{G2}, 
$p=0.20$, \textbf{H3a-SIM rejected}, whereas in the \textit{random variables} lesson, a significant difference was found, $p<0.05$, \textbf{H3a-COM supported}.
% This might be due to the \textit{fundamental graphs} lesson's simpler course content and limited test questions.
This suggests that the \textit{full version} of \toolName{} is more effective in aiding learners to understand complex courses.
For simpler courses, the abundance of visual designs provided by the \textit{full version} may be excessive. Learners might rely solely on the timeline-based course structure for comprehension. Nevertheless, the \textit{full version} still demonstrates an advantage in terms of test scores.}
We interviewed \textit{G1}, and they agreed that \toolName{} effectively helped them understand the overall course structure. The visualizations and interactions enabled them to perform well on structured topics (FG: 4.4, RV: 4.5). P3 noted that: ``Automatically constructing concept maps saves time and effort for taking notes, and I can easily discover course organization and knowledge dissemination." We also found that participants tended to do repetitive exploration of important concepts during learning, such as P21, P22, P24, and P27 consolidating concepts after watching the video in preparation for the test.}
% This viewpoint was supported by P1 ``Thanks to \toolName{}, I can find important concepts faster and navigate quickly."}

\textit{\uline{\revise{Time Costs.}}}
The difference in learning time highlights \toolName{}'s ability to accelerate learning while maintaining comprehension. Compared to \textit{G3} (FG: 9m03s, RV: 10m44s),~\textit{G1} (FG: 5m42s, RV: 7m13s) allowed faster learning, as did \textit{G2} (FG: 7m14s, RV: 9m27s),
\minorRevision{(Fig.~\ref{FigSig}, \textbf{H1b-SIM supported}, \textbf{H1b-COM supported}).}
\minorRevision{The Bonferroni post-hoc test also detected a significant reduction in time costs for \textit{G1} compared to \textit{G2} ($p<0.05$, \textbf{H3b-SIM supported}; $p<0.05$, \textbf{H3b-COM supported}), indicating the \textit{full version} of \toolName{} can effectively facilitate learning.}
P8 stated, ``I often watch videos at double speed, this system helps me to quickly ignore the less important parts and focus on the important concepts". P15 and P26 shared this habit and expressed anticipation for using the \textit{full version} of \toolName{}. There is an outlier in time spent, P6, who spent more time on the \textit{random variables} lesson due to a perceived concept map error. He used the edit mode to fix the concept map dynamically. P6 commented, ``The edit mode allowed me to modify the concept map, which took some time, but I gained a longer-lasting memory". P1 and P3 confirmed the issue but felt it did not have a significant impact on their understanding. ``When I find an error. I go straight to the original video by navigating to it to find the correct organization, it doesn't take me much time" (P1). The math majors already possessed prior knowledge of the ``random variable" lesson, so they spent less time and obtained relatively high test scores. Nonetheless, \toolName{} also provided effective assistance. As P9 commented, ``The system's concept maps allowed me to quickly review the concepts and enhanced understanding to some extent."

\textit{\uline{\revise{CLS Results.}}}
\revise{The right side of Fig.~\ref{Fig8} shows the results of the Cognitive Load Survey (CLS), which partially reflected the cognitive impact brought by \toolName{}. The intrinsic load (IL) corresponds to the difficulty of the learning material, and its numerical representation remains constant. To minimize subjective judgment errors, we aligned the results based on the IL. Overall, participants positively evaluated the reduced cognitive load in all versions. Specifically, \textit{G1} (FG: 1.2, RV: 1.5) and \textit{G2} (FG: 1.9, RV: 2.1) outperformed \textit{G3} (FG: 2.2, RV: 2.4) due to the intuitive representation of course structure,
\minorRevision{(Fig.~\ref{FigSig}, \textbf{H2a-SIM supported}, \textbf{H2a-COM supported})}, which reduced the extraneous load (EL) from the tool and reading intervention.
\minorRevision{We also conducted the Bonferroni post-hoc test for EL, and the results indicate a significant reduction in the cognitive load for \textit{G1} compared to the \textit{G2} ($p<0.05$, \textbf{H4a-SIM supported}; $p<0.05$, \textbf{H4a-COM supported}).}
This saved effort in organizing concepts, freeing more working memory capacity for generative tasks. As a result, \textit{G1} (FG: 6.7, RV: 7.4) and \textit{G2} (FG: 7.1, RV: 7.8) generated less germane load (GL) than \textit{G3} (FG: 7.4, RV: 8), 
\minorRevision{
(Fig.~\ref{FigSig}, \textbf{H2b-SIM supported}, \textbf{H2b-COM supported}), which refers to students' self-perceived learning.
Although \textit{G1} had a lower cognitive load on the mean score, a Bonferroni post-hoc test indicated that there was no significant difference found in the \textit{fundamental graphs} lesson compared to \textit{G2}, $p=0.12$, \textbf{H4b-SIM rejected}. In the \textit{random variables} lesson, \textit{G1} was significantly different from \textit{G2}, $p<0.05$, \textbf{H4b-COM rejected}.
% This indicates that \toolName{} is more effective in reducing cognitive load in complex courses, a benefit less evident in simpler courses.
This indicates that \toolName{} is more effective in reducing cognitive load for complex courses.
In the case of simpler courses, where the cognitive load is inherently low, the richer externalization of knowledge does not significantly reduce the cognitive burden on learners. However, in terms of overall performance, the \textit{full version} still holds an advantage over the \textit{primitive version}.
}}
% This is ultimately reflected in the fact that the \textit{full version} (FG: 4.6, RV: 4.1) and the \textit{primitive version} (FG: 4.2, RV: 3.7) can help comprehensibility problems better compared to the \textit{original video} (FG: 3.8, RV: 3.4).

\begin{figure}[t]
    \centering
    \includegraphics[width=\linewidth]{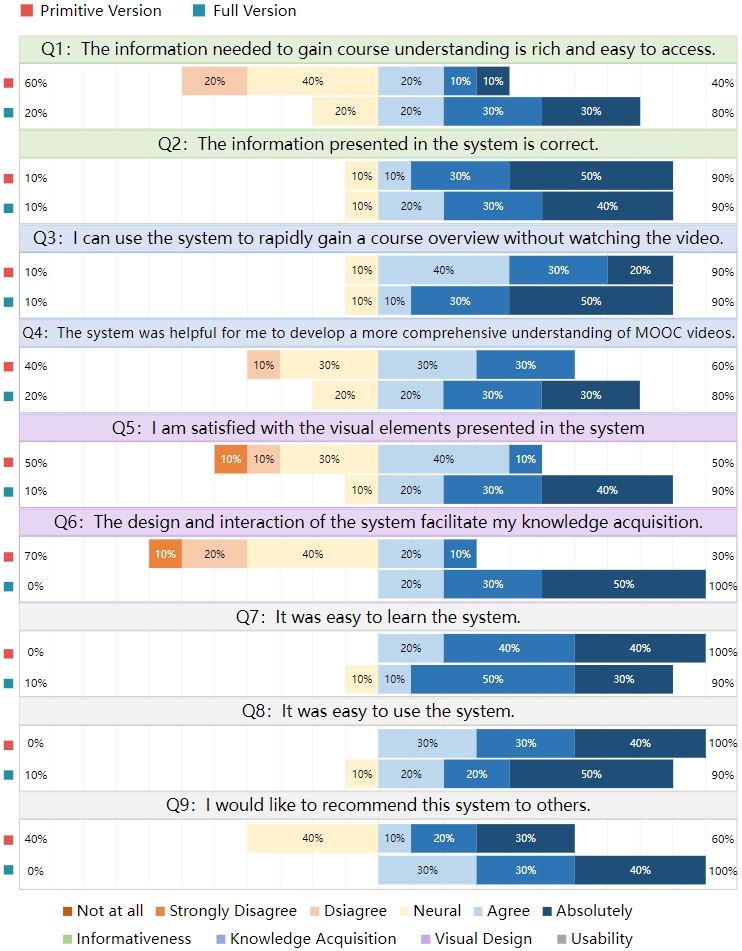}
    \caption{\minorRevision{Comparison of participants' agreements on \toolName{}'s full and primitive version from the post-study questionnaire. \textit{Agree} scores to the right, \textit{Neutral} or~\textit{Disagree} to the left. No participant selected \textit{Not at all}.}}
    \label{Fig9}
    \vspace{-4mm}
\end{figure}

\textbf{\revise{Questionnaire.}}
\revise{As shown in Fig.~\ref{Fig9}, participants provided positive feedback on \toolName{}. The \textit{full version} was preferred for its informativeness and knowledge acquisition support. It scored slightly lower in system information correctness due to handling complex content. The \textit{full version} was also favored for its visual design and usability, making it more appealing and confidence-inspiring. Participants, even those with extensive MOOC experience, preferred and recommended the \textit{full version} of \toolName{}.}

\section{Discussion}

This section explores the limitations of the systems, design reflections learned from this process, and implications for the education domain.

\textbf{System Limitations.}
\revise{We discuss several} limitations of the initial \toolName{}. 
First, although we included many important elements, there are still others that could be added to reduce cognitive loads, such as changes in voice \revise{modulation} and learner interactions. Furthermore, to complete the knowledge network, various relationship types must be considered. 
To achieve this, we propose organizing relations between concepts based on their logical and narrative relationships~\cite{wolfinger2008plug}. 
In this case, we can construct a more robust network of concept maps that reveal the course structure and details between concepts~\cite{shahaf2012trains}.
Second, in our data \revise{processing}, we did not use additional information to extract more precise results. We have only divided the \revise{course style} into three categories, and this approach exhibits irregularity due to the instability of structural features and the strong dependence on teachers' habits. To address this issue, we can classify handwritten materials based on their temporal order. Additionally, we have conducted detailed content analysis on slides, but rich graphical content has not been analyzed yet. This can be improved through emerging CV algorithms~\cite{dong2021visual} to extract concepts and relationships from charts and images.
Ultimately, in the concept relationship extraction phase, we used the large language model GPT-3.5, and thanks to GPT-3.5's self-learning and self-optimization capabilities, it can continuously improve its prediction and generation abilities. Furthermore, with the release of the latest model, GPT-4, these issues will gradually be resolved in the future.

\textbf{Design Reflections.}
In the current \toolName{} design, we focus on encoding analyzed data in concept maps but lack support for raw data. Through communication, learners have expressed a need to access raw data to aid in understanding. Therefore, in future versions, we plan to incorporate multi-word phrases or sampled references to create keyword clouds to support learners' understanding of concepts.
Additionally, we aim to create a concise design that effectively summarizes and visualizes MOOC content for learners to browse or review, using a modular architecture built on a plugin-based approach as demonstrated in previous work~\cite{wolfinger2008plug,nobarany2014designing}. Thus, we plan to offer a plugin version that intentionally omits the standard data flow in visual analysis systems, especially access to raw videos, and can be easily integrated into fully-featured systems.
Finally, the IH four-step learning stage model~\cite{haring1978fourth} indicates that the final stage of learning should be the incorporation of knowledge into application. To facilitate this, \revise{future work in visual analytics} could focus more on enhancing the ability of systems to analyze user-generated insights, thereby better assisting users in achieving their visualization goals. 
\revise{This is particularly important in cases where long-term memory is required to retain the cognitive output.}

\textbf{Implications for the Education Domain.}
After a comprehensive review of learning content analysis and discussions with experts, we have further summarized three potential directions for \toolName{} in education.
The first direction is to explore the generalization of \toolName{}. As MOOCs are commonly used in education, we believe that studying them systematically can guide other educational methods. For example, we can apply similar systems to analyze learning in classrooms, and extract concepts based on interactions between teachers and students. Also, we can apply this to other rich knowledge-based sequential data like speeches and debates. Analyzing and extracting their organization and content can help learners understand the general theme, speech mode, and arguments of both sides.
% Therefore, \toolName{} is versatile and adaptable to various application fields.
The second direction is to leverage student data to evaluate course design and provide guidance for a redesign. By \revise{recording user performance while using the system}, teachers can identify common learning challenges and adjust course content accordingly, while learners gain insight into their own knowledge acquisition. Moreover, based on learners' feedback, we can share this data among teachers, which can facilitate effective course organization and course structuring.
% By conducting comparative analyses of various course organizations and course styles to determine their impact on learners' conceptual understanding across a larger number of courses and learners, we can also generate more valuable insights for the education field.
The third research direction is to build a comprehensive concept network. This entails analyzing videos in sequential order and leveraging the organizational patterns between courses to construct a holistic concept network. By utilizing richer metadata such as forums, Q\&As, and assignments, more diverse relationships can be extracted, enabling learners to build a course-level mental model.

\section{Conclusion}
We present \toolName{} for analyzing MOOC videos. It incorporates a number of new visual \revise{designs} and multiple interactive views \revise{to present concepts through a thread metaphor, aiding learners in obtaining a better understanding of MOOC videos.} It supports a rich set of interactions, allowing for flexible visual exploration between views. We also introduce data structures and methods for extracting concepts and their relationships from MOOC videos, as well as \revise{design requirements} summarized from interviews with domain experts. \revise{A quantitative study, two case studies, and a user study} demonstrate the effectiveness and usability of the \toolName{}.


% \appendix % You can use the `hideappendix` class option to skip everything after \appendix
\vspace{-6mm}
\begin{IEEEbiography}[{\includegraphics[width=1.0in,height=1.25in,clip,keepaspectratio]{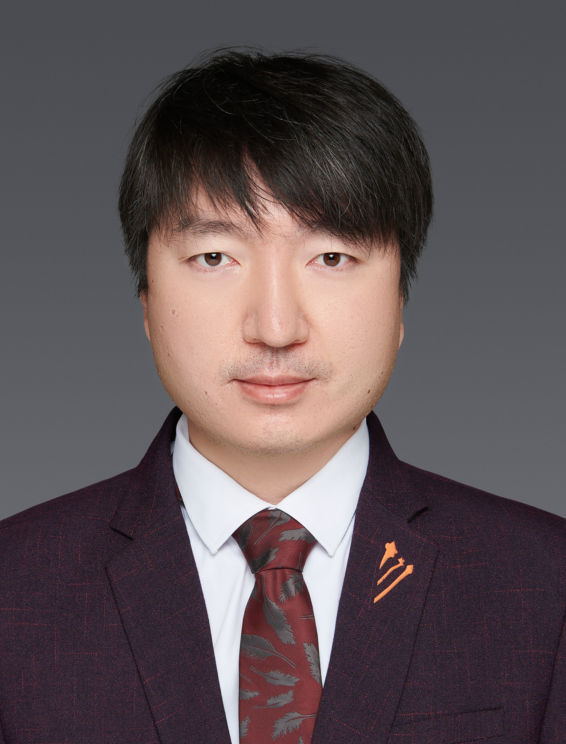}}]{Zhiguang Zhou} is currently a professor in School of Media and Design and serves as the director of Digital Media Technology Research Institute at Hangzhou Dianzi University. His research interests include data visualization, visual analytics and online learning data analysis. He received his Ph.D. in Computer Science from the state key Laboratory of CAD\&CG in Zhejiang University. 
\end{IEEEbiography}

\begin{IEEEbiography}[{\includegraphics[width=1.0in,height=1.25in,clip,keepaspectratio]{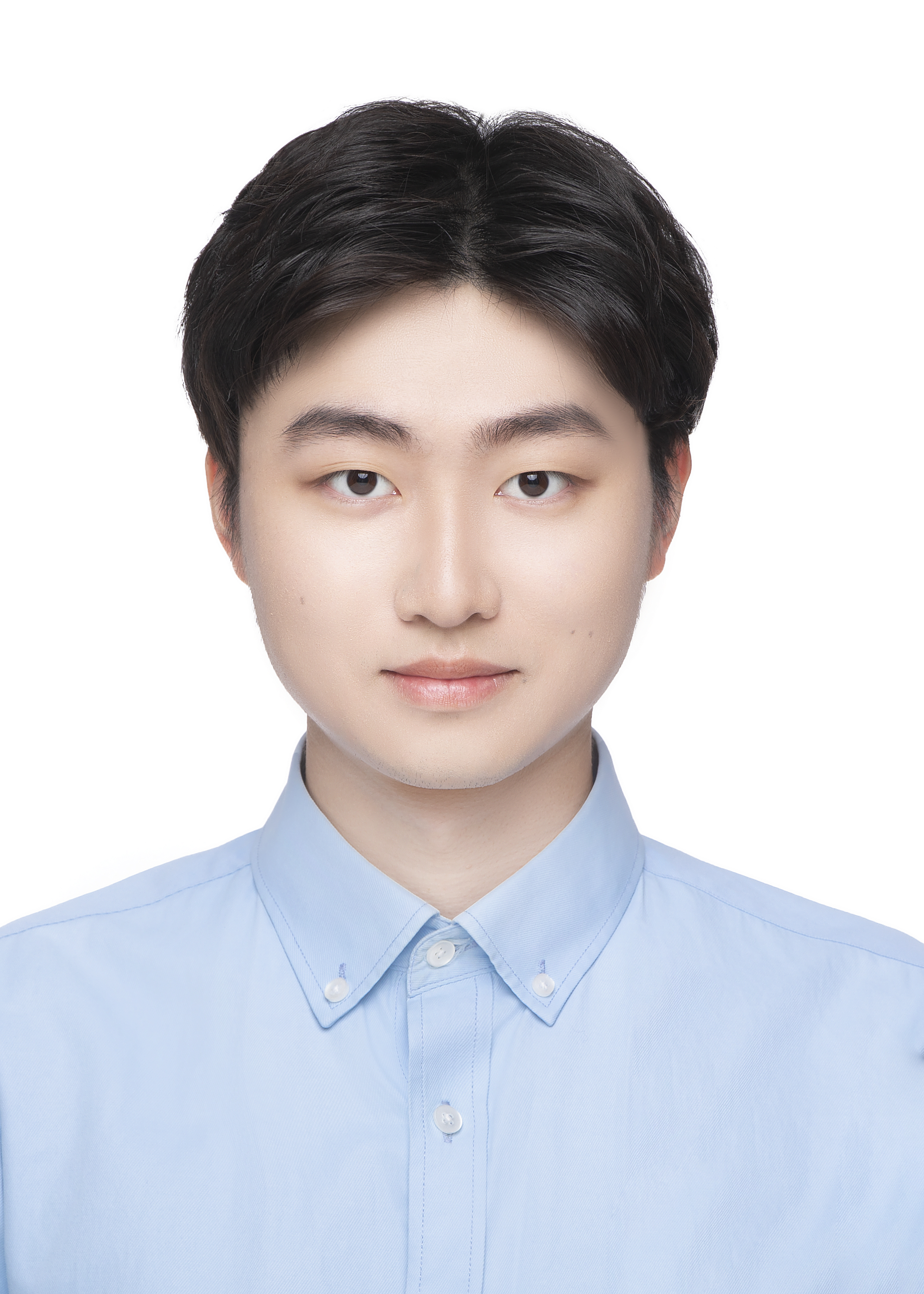}}]{Li Ye} is currently working toward the postgraduate degree with the School of Media and Design, Hangzhou Dianzi University, Hangzhou, China. His interests include visual analytics and visual storytelling.
\end{IEEEbiography}

\begin{IEEEbiography}[{\includegraphics[width=1.0in,height=1.25in,clip,keepaspectratio]{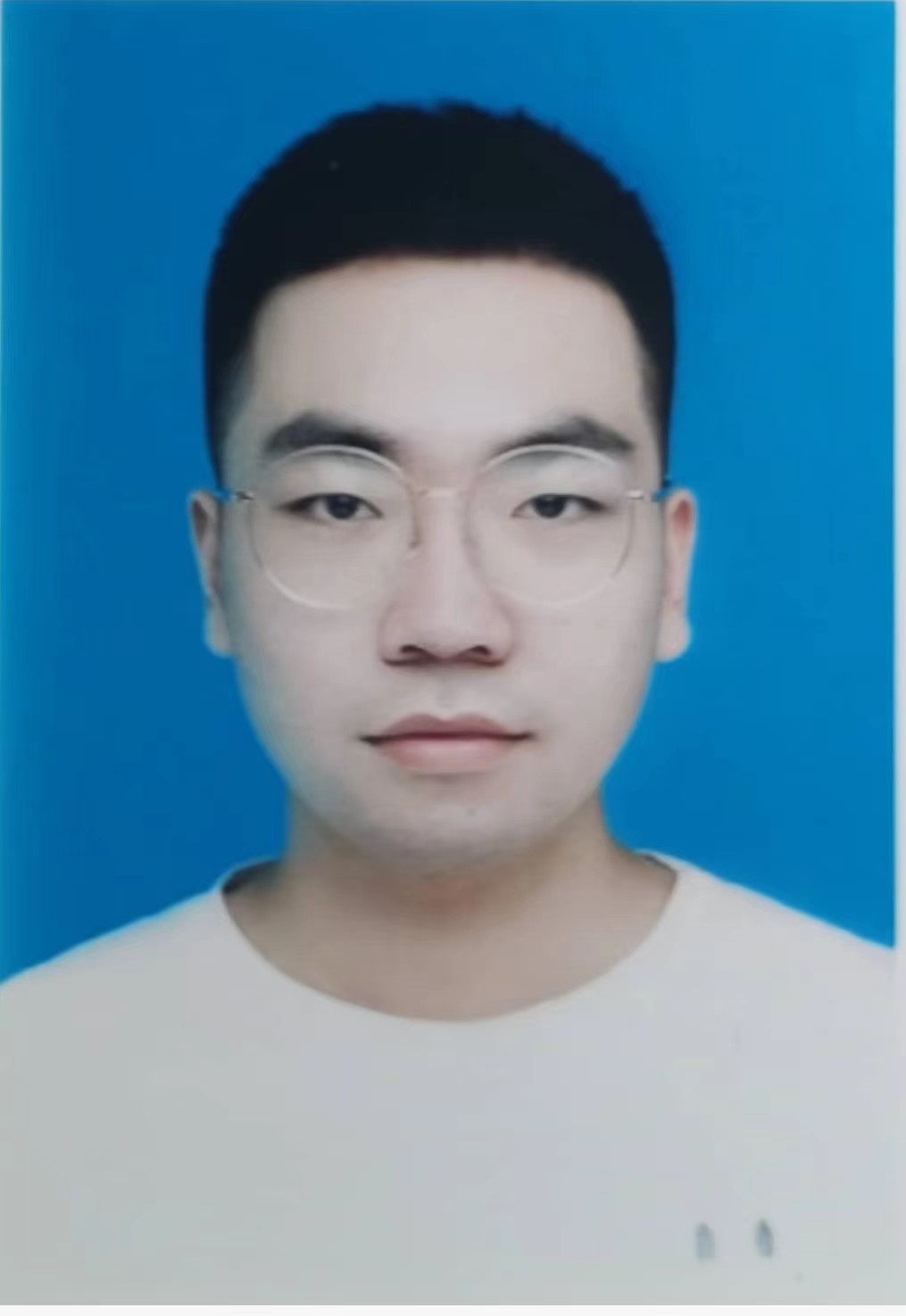}}]{Lihong Cai} is currently a MPhil student at Hangzhou Dianzi University. His research interests include data visualization and human-computer interaction.
\end{IEEEbiography}

\begin{IEEEbiography}[{\includegraphics[width=1.0in,height=1.25in,clip,keepaspectratio]{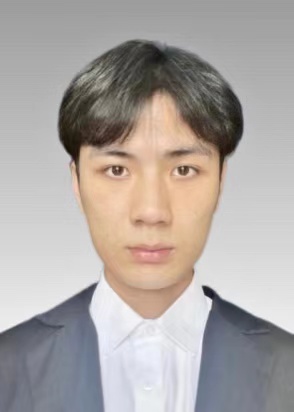}}]{Lei Wang} is currently a MPhil student at Hangzhou Dianzi University. His research interests include information visualization and visual analytics.
\end{IEEEbiography}

\begin{IEEEbiography}[{\includegraphics[width=1.0in,height=1.25in,clip,keepaspectratio]{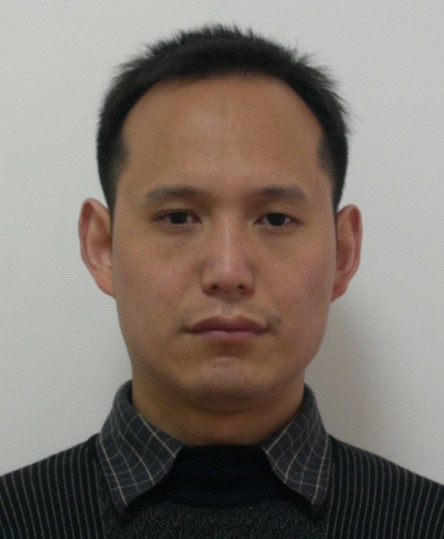}}]{Yigang Wang} received the MS and PhD degrees in applied mathematics from Zhejiang University, Hangzhou, China. He is currently a professor with the School of Media and Design, Hangzhou Dianzi University, Hangzhou, China. His interests include image processing, computer vision, pattern recognition, and computer graphics.
\end{IEEEbiography}

\begin{IEEEbiography}[{\includegraphics[width=1.0in,height=1.25in,clip,keepaspectratio]{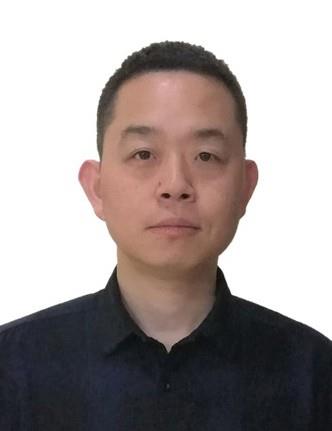}}]{Yongheng Wang} received the Ph.D. degree in computer science and technology from the National University of Defense Technology, Changsha, China, in 2006., He is currently a Research Specialist with the Research Center of Big Data Intelligence, Zhejiang Lab, Hangzhou, China. His research interest covers big data analysis, machine learning, computer simulation, and intelligent decision making.
\end{IEEEbiography}

\begin{IEEEbiography}[{\includegraphics[width=1.0in,height=1.25in,clip,keepaspectratio]{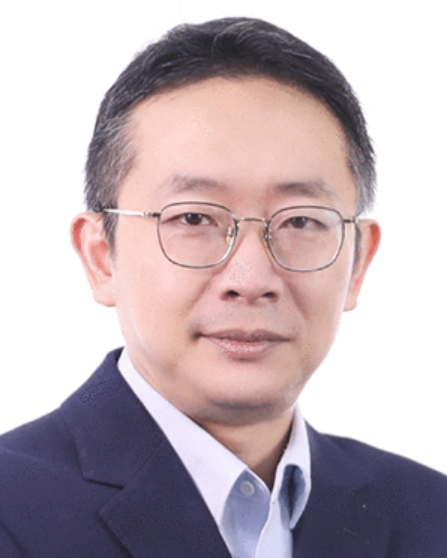}}]{Wei Chen} is a professor with the State Key Lab of CAD\&CG, Zhejiang University. His research interests include visualization and visual analysis. He has published more than 70 IEEE/ACM Transactions and IEEE VIS papers. He actively served as guest or associate editors of the ACM Transactions on Intelligent System and Technology, IEEE Computer Graphics and Applications and Journal of Visualization.
\end{IEEEbiography}

\begin{IEEEbiography}[{\includegraphics[width=1.0in,height=1.25in,clip,keepaspectratio]{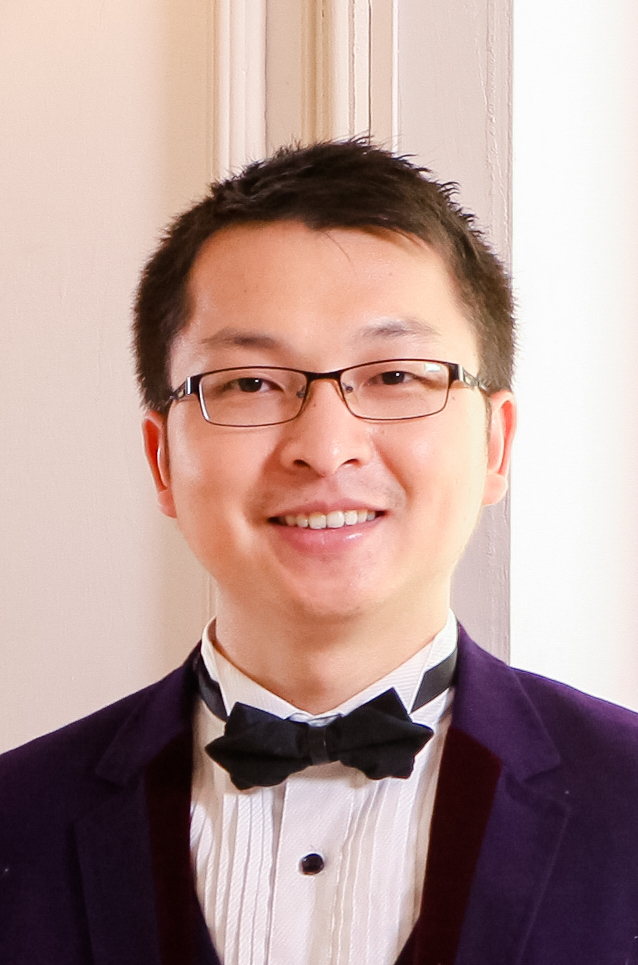}}]{Yong Wang} is currently an assistant professor in School of Computing and Information Systems at Singapore Management University. His research interests include data visualization, visual analytics and explainable machine learning. He obtained his Ph.D. in Computer Science from Hong Kong University of Science and Technology. He received his B.E. and M.E. from Harbin Institute of Technology and Huazhong University of Science and Technology, respectively. For more details, please refer to http://yong-wang.org.
\end{IEEEbiography}

\end{document}